\documentclass[fleqn,usenatbib]{mnras}
\usepackage{newtxtext,newtxmath}
\usepackage[T1]{fontenc}
\DeclareRobustCommand{\VAN}[3]{#2}
\let\VANthebibliography\thebibliography
\def\thebibliography{\DeclareRobustCommand{\VAN}[3]{##3}\VANthebibliography}
\usepackage{graphicx}	
\usepackage{amsmath}	
\usepackage{multicol}        
\usepackage{bm}		
\usepackage{pdflscape}	
\usepackage{hyperref}
\usepackage{siunitx} 
\usepackage{gensymb}

\title{Low-resolution Transit Spectroscopy of Three Hot Jupiters Using the 2m Himalayan Chandra Telescope}
\author[Athira Unni et al.]{
Athira Unni,$^{1,2}$\thanks{E-mail: athira.exo@gmail.com} Thirupathi Sivarani,$^{1}$  Jayesh Goyal,$^{3}$ Yogesh C. Joshi,$^{2}$ Apurva V. Oza $^{4}$
\newauthor
Ravinder K Banyal$^{1}$
\\
$^{1}$Indian Institute of Astrophysics, Koramangala 2nd Block, Bangalore, India\\
$^{2}$ Department of Physics and Astronomy, University of California, Irvine, USA\\
$^{3}$ National Institute of Science Education and Research, Bhubaneswar, Odisha, India\\
$^{4}$ Aryabhatta Research Institute of Observational Sciences, Nainital, Uttarakhand, India\\
$^{5}$ Jet Propulsion Laboratory, California Institute of Technology, Pasadena, USA \\}

\date{Accepted XXX. Received aYYY; in original form ZZZ}
\pubyear{2015}

\begin{document}
\label{firstpage}
\pagerange{\pageref{firstpage}--\pageref{lastpage}}
\maketitle
\begin{abstract}

Here, we present the low-resolution transmission spectroscopy of three giant planets using the Himalayan Faint Object Spectrograph Camera (HFOSC) on the 2m Himalayan Chandra Telescope (HCT) in Hanle, India. It is the first application of transmission spectroscopy with HCT. This study presents results from a single transit, each for three planets: HAT-P-1b, KELT-18b and WASP-127b. The selection of suitable reference stars assisted in accurately tracking slit losses for the long cadence observations that are needed to achieve the required Signal to Noise Ratio (SNR). We employ the Common Mode Correction (CMC) technique, utilizing a white light transit curve to minimize time dependent systematic errors. The observed spectra for WASP-127b and HAT-P-1b agree with previous low-resolution transit spectroscopic observations using other observing facilities. We confirm the presence of Rayleigh scattering in the atmosphere of WASP-127b. In addition, we provide the first low-resolution transmission spectrum for KELT-18b. Modeling the exoplanet atmosphere with HFOSC and available IR observations from HST and SPITZER for WASP-127b and HAT-P-1b shows that HFOSC can be an alternative optical instrument to use in conjunction with IR observations to constrain the atmospheric parameters better.

\end{abstract}


\section{Introduction}
Exoplanet transit observation using low-resolution spectroscopy is achieved by taking a time series of spectrophotometric observations during transit and measuring the transit depth as a function of wavelength. The observed variation of transit depth with wavelength is due to various sources of opacities set by the absorption and scattering properties of gases and aerosols near the planet's terminator \citep{Seager_2000, Brown_2001}. Early progress in understanding the exoplanet atmosphere started with low-resolution transit spectroscopic observations from space-based telescopes, free from Earth's telluric features \citep{Charbonneau_2002, Bean2010, Snellen2010, Sing2016}.  These observations have constrained the atmospheric characteristics of several exoplanets, detecting new elements and understanding the thermal structure. \citet{Sing2016} performed a comprehensive analysis by examining the transmission spectra of ten hot Jupiters using HST and Spitzer and found that the weakened spectral signatures in the observed spectra are due to the presence of clouds and hazes present in the atmosphere rather than the depletion of primordial water. This work also emphasizes the importance of population study of the exoplanet atmosphere to understand their formation and evolution processes.

Ground-based multi-object and long-slit observations enabled accurate differential spectrophotometry by simultaneously observing a reference star to track variable slit loss. The transiting exoplanet survey using Gemini/GMOS \citep{Hook_2004} observed several exoplanet atmospheres and showed that a single transit observation is capable of successfully constraining the atmospheric parameters  \citep{Todorov_2019,Wilson_2021,Panwar_2022}. VLT/FORS2 \citep{Appenzeller_1998} observed the upper atmospheric hazes and the presence of alkali elements in many exoplanetary atmospheres \citep{Nikolov_2016,Lendl_2016,Spyratos_2021}, including the super-Earth GJ 1214b by \citet{Bean_2010}. Through the ACCESS Survey, Magellan/IMACS \citet{Dressler_2011} has observed the atmosphere of several planets, ranging from sub-neptunes to inflated hot jupiters, providing insights into their atmospheric structure, absorbers, and stellar activity \citep{Bean_2011,Rackham_2017,Weaver_2020}. Along with characterizing diverse exoplanets, GTC/OSIRIS were able to detect potassium in the atmosphere of many planets using narrow band spectrophotometry \citep{sing_2012,Murgas_2014,wilson_2015,Murgas_2019, Chen_2020}. Similarly, LBT/MODS \citep{Mallonn_2016,Yan_2020} also effectively contributed to ground-based low-resolution transit spectroscopic observations. 

Due to the larger scale height, hot Jupiters and Saturns are the preferred candidates to observe from the ground \citep{Seager_1998}. Strong absorption lines of alkali metals such as Na doublets (589 nm) \citep{Charbonneau_2002,Snellen_2008,Nikolov_2016,Sing2016,Nikolov_2018}, K I doublets (767 nm) \citep{sing_2011, Nikolov_2015, Sing_2015, Sing2016, Sedaghati_2016, Chen_2018}, and Li (670.7 nm) \cite{Chen_2018} are most commonly detected features in the atmosphere of hot Jupiters and Saturns because of the large absorption cross section. Rayleigh scattering in the bluer wavelength is also detectable in the visible part of the spectra \citep{jordan_2013, Sing_2015, Palle_2017, Murgas_2019, May_2020}. 

A non-detection of molecular and/or atomic species \citep{Gibson_2013, Sing_2015, Nortmann_2016, Weaver_2020} are also important. High-altitude clouds or hazes in the atmosphere of a gaseous planet can suppress the observed atomic and molecular signals in the spectra. Observations from different instruments or repeated observations for the same planet over multiple transits have sometimes revealed different results and stellar disk in-homogeneity could be a possible cause \citep{Rackham_2018}. Some studies using 8m-class telescopes \citep{Todorov_2019,Jiang_2023} show that it can be difficult to reproduce the detection from a smaller space-based telescopes. 
Also, accurate telluric correction at low resolution is challenging. Hence, ground-based transit observations are often not photon noise limited but affected by systematic errors. Repeated observations of the same planet using various instruments during various epochs will help determine the robustness of the results \citep{Palle_2017, Todorov_2019, May_2020}.

Ground-based low-resolution transit spectroscopic observations suffer from contamination of telluric lines and variable slit losses, comparable to the amplitude of variations from the planetary atmosphere. Simultaneous multi-object observation can minimize these errors. Differential photometric corrections work well when the brightness and spectral type of comparison stars are similar to that of the target stars \citep{Nascimbeni_2013}. Here we present low-resolution transit spectroscopic observations of three hot Jupiters for the first time using the 2m Himalayan Chandra Telescope. This is also the first successful transmission spectroscopic observation ever made from HCT. The first object, HAT-P-1A is a visual binary star with a twin companion.  The planet host in the system, HAT-P-1A is a G-type star with an inflated Jupiter-sized planet of radius $1.2R_{J}$, mass of $0.52M_{J}$, and an equilibrium temperature of 1322 K \citep{hatp1_detection, Nikolov_2014}. \citet{Nikolov_2014} detected sodium (589 nm) in the atmosphere of HAT-P-1b using HST/STIS for the first time. Subsequently, \citet{wilson_2015} successfully identified potassium using GTC/OSIRIS. Building upon this, \citet{Chen_2022} detected sodium and potassium in the atmosphere of HAT-P-1b using P200/DBSP telescope. However, the 8m Gemini/GMOS \citep{Todorov_2019} observation could not confirm sodium detection. Furthermore, \citet{Montalto_2015} observed only a blue edge of the sodium broad absorption wing using DOLORES/TNG observations.

The second object, WASP-127, is the puffiest, low-density planet around a G5 star, also part of a binary system, and hosts a planet with an equilibrium temperature of 1400 K, radius of $1.37R_{J}$ and mass of $0.18M_{J}$ in the transition region of Jupiter to Neptune mass \citep{Lam2017}. HST and Spitzer observations by \citet{skaf_2020,Spake_2021} showed feature-rich transit spectra with absorption from sodium and wavelength-dependent Rayleigh scattering in the optical. \citet{Palle_2017} studied the atmosphere of WASP-127b using 2.5m NOT/ALFOSC and detected a strong Rayleigh scattering slope and a hint of sodium detection. The third target, KELT-18 is a single star hosting a massive planet of $R_{p}=1.57R_{J}$ and $M_{p}=1.18M_{J}$ \citep{kelt18_detection}, and this work presents low-resolution transit spectroscopy for KELT-18b for the first time. In the case of HAT-P-1A and WASP-127, we used the binary companion star as the reference star, and for KELT-18, we used the nearby bright field star as a reference star. None of the three reference stars are known to be variable. Our attempt of demonstrating the feasibility of a 2m class telescope will encourage the community to use more similar telescopes for observing the exoplanet atmosphere.

The paper is organized as follows. Section \ref{obs} covers the observations, while Section \ref{method} explains the data reduction and analysis. Section \ref{model} presents details of the modeling of the observed data and the results obtained. Finally, the discussion and conclusions are presented in Section \ref{discussion}. Additional figures supporting this study can be found in Appendix \ref{appendix}.

\section{Observations} \label{obs}
Observations are obtained from the Indian Astronomical Observatory (IAO) at Hanle, the Ladakh region of the Himalayas, operated by the Indian Institute of Astrophysics. We use HFOSC \footnote{\url{https://www.iiap.res.in/?q=iao_2m_hfosc}}, a low-resolution slit spectrograph uses different grisms and slits of varying widths and lengths, mounted on the 2m Himalayan Chandra Telescope. HFOSC has a field of view of $10\arcmin \times 10\arcmin$ and a $2048 \times 4096$ SITe detector with a pixel scale of
$0.296\arcsec$. We used grism-8 and grism-7 to cover the wavelength range of 3800 {\AA} to 6500 {\AA}  and 5800 {\AA} to 8350 {\AA}, respectively.
We performed differential spectrophotometry with the longest and widest available slit ($11^{\arcmin} \times 15.41^{\arcsec}$) to minimize the time variable slit loss and maximize FOV with a resolution of 1180 (grism-7, at 5900 {\AA}) and 1920 (grism-8, at 7800 {\AA}). Only one of the grism settings is used for a given transit; hence, the entire optical wavelength range is not covered simultaneously. Choosing a reference star is also essential to minimize differential slit loss between the target and reference as a result of atmospheric dispersion, seeing, and other environmental and instrument effects. A good reference star could be a visual twin binary companion or a field star of similar spectral type and brightness to the target star. Due to the FOV, we are also limited to a maximum separation of $11\arcmin$ between the exoplanet host and a reference star. During observation, we aligned the slit along the target star and the reference star by changing the position angle of the instrument cube in the cassegrain unit so that simultaneous observations of the target and reference are possible. The details of the observations and objects are given in Table \ref{table1} and Table \ref{table2}, respectively. 

\begin{table*}
\begin{center}
\caption{{\bf Observational details: } Name of the target, Date of Observation (DoO), Peak Signal to Noise Ratio per exposure (SNR), Exposure time (Exp.time), Number of exposures (No.Exp), Parallactic angle (PA) in degrees, Position Angle (PoA) in degrees, airmass at the beginning and end of the observation, grism used and the corresponding wavelength coverage, and the name of the reference star used are listed.}
\begin{tabular}{crrrrrrrrrr}
\hline\hline
Object&DoO&SNR&Exp.time&No.Exp&PA&PoA (E of N)&Airmass& Grism &Wavelength Range&Reference Star\\
\hline
WASP-127 &$09-03-2019$&$909$& 420s&62& 339.19&78.2&1.99-1.99& 7&3800 - 6500 {\AA}&TYC 4916-897-1\\
HAT-P-1A &$02-07-2018$&$944$&$180s$&57&272.45&-99.5&1.55-1.02&7&3800 - 6500 {\AA}&HAT-P-1B\\
KELT-18&$10-06-2019$&$1075$&$420s$&55&125.34&2.0&1.15-1.74&8&5800 - 8350 {\AA}&TYC 3865-1339-1 \\
\hline
\label{table1}
\end{tabular}
\end{center}
\end{table*}

\begin{table*}
\begin{center}
\caption{{\bf Stellar parameters:} Name of the target star and the reference star (Object), visual magnitude (Vmag), effective temperature ($T_{eff}$) in Kelvin, the logarithm of gravity ($\log{g}$), and metalicity ([Fe$/$H]) and separation between the target and the reference star are given below. }
\begin{tabular}{cccccrrrrrr}
\hline\hline 
Object&Vmag &$T_{eff}$& $\log{g}$ &[Fe$/$H]& Separation \\
\hline
WASP-127 &$10.20$&$5750\pm100$&$3.90\pm0.1$& $-0.18\pm0.06$&41.8\arcsec\\
TYC 4916-897-1&$11.0$&$5733$&$4.480$&...&...\\
HAT-P-1A&$10.4$ &$6047\pm56$&$4.13\pm0.10$&$0.12\pm0.05$&11.2\arcsec\\
HAT-P-1B&$9.87$&$5975\pm45$&$4.45\pm0.06$&$0.13\pm0.02$&...\\
KELT-18&$10.4$&$6670\pm120$&$4.034\pm0.083$&$0.08\pm0.13$&5.79\arcmin\\
TYC 3865-1339-1&$9.92$&$6787$&$3.976$&...&...\\
\hline
\end{tabular}
\label{table2}
\end{center}
\end{table*}

\section{Data reduction} \label{method}
\begin{figure*}
\centering
\includegraphics[width=0.65\columnwidth]{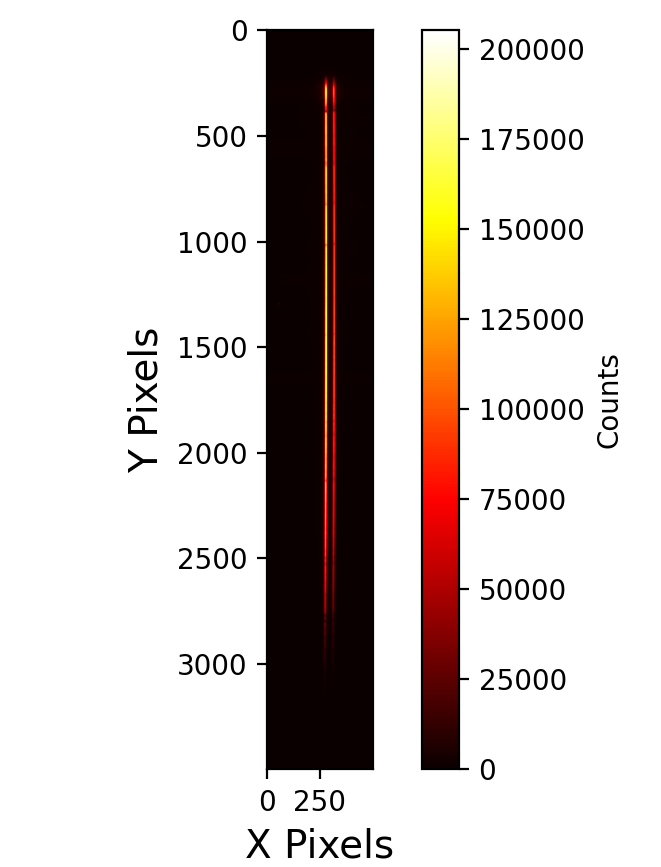}
\includegraphics[width=0.65\columnwidth]{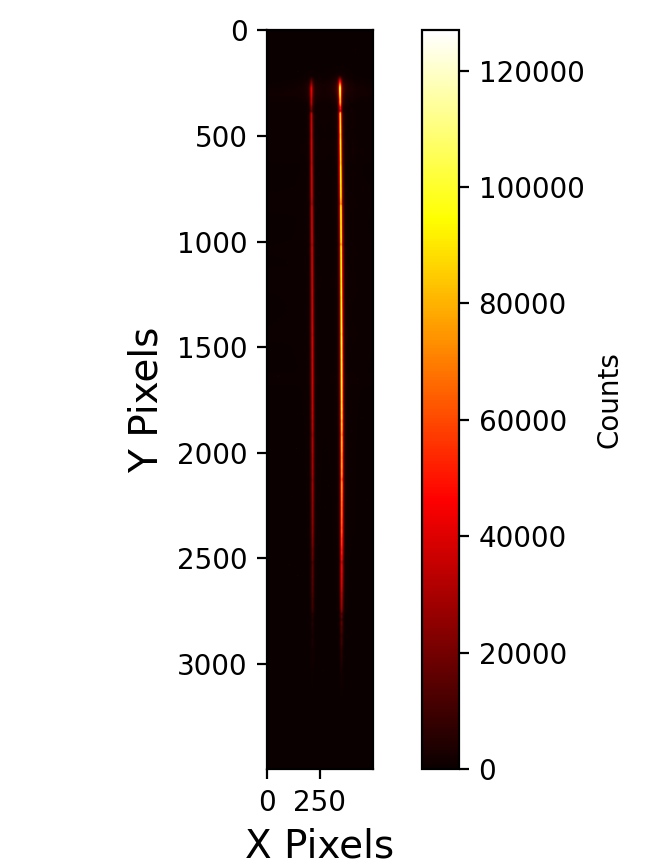}
\includegraphics[width=0.65\columnwidth]{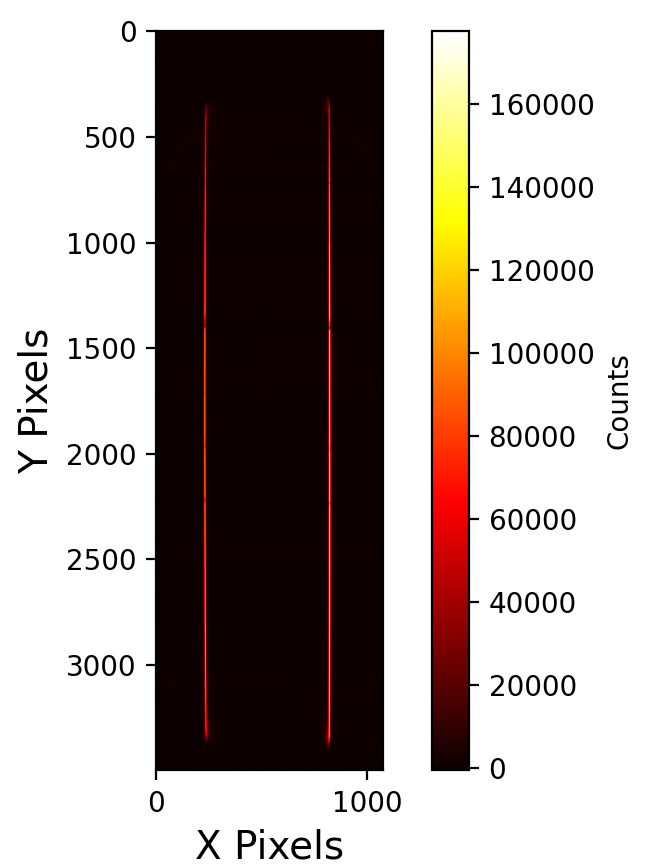}
\caption{HCT/HFOSC single exposure CCD images of HAT-P-1A, WASP-127, and KELT-18, along with the reference star from left to right. The horizontal and vertical axis correspond to the position along the slit and the wavelength, respectively. }\label{stars_ccd_image}
\end{figure*}

\begin{figure*}
\centering
\includegraphics[width=0.7\columnwidth]{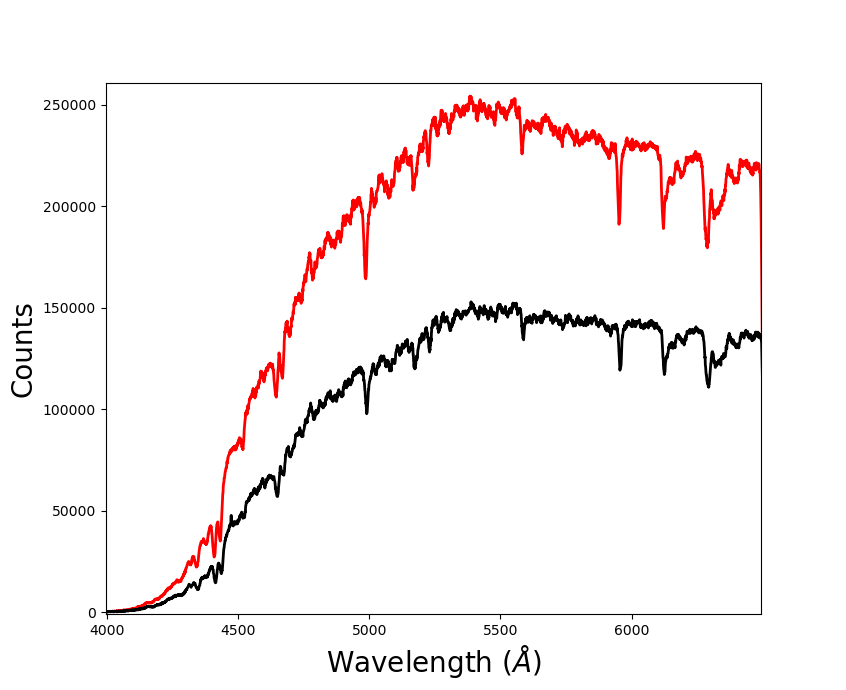}
\includegraphics[width=0.67\columnwidth]{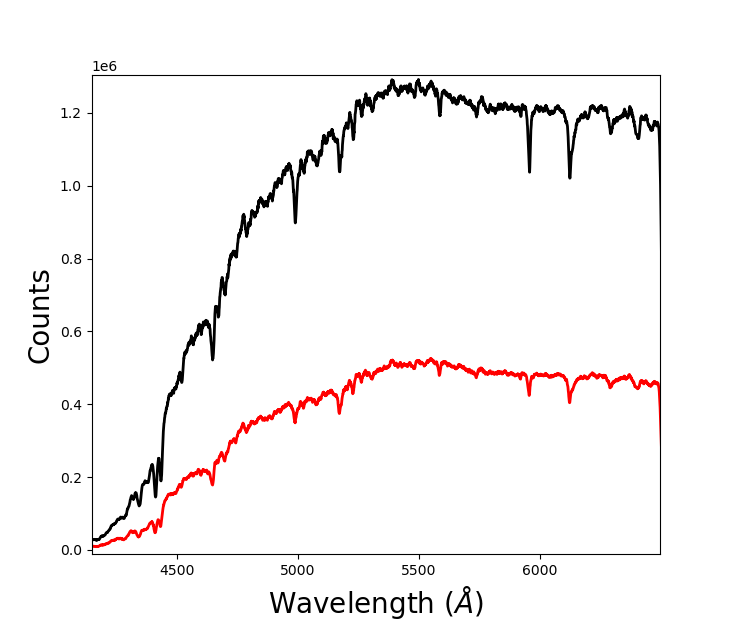}
\includegraphics[width=0.675\columnwidth]{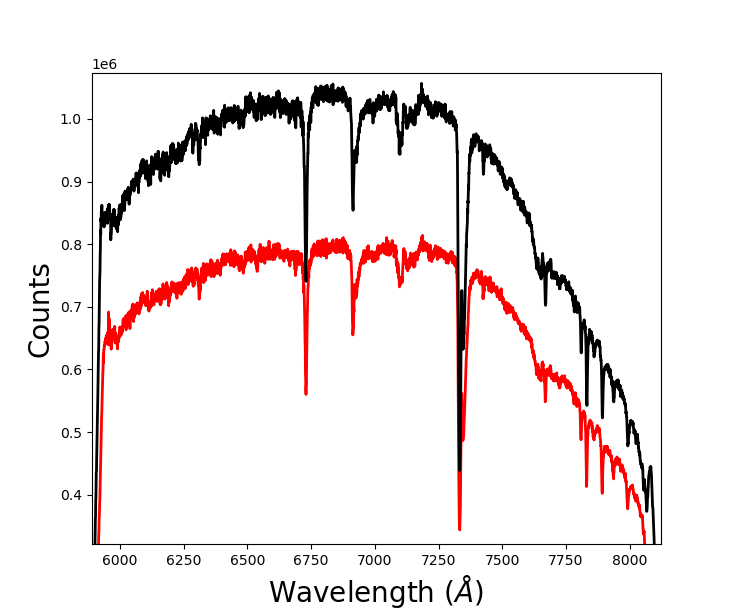}
\caption{ Single exposure, wavelength calibrated spectra of HAT-P-1A (Grism 7), WASP-127 (Grism 7), and KELT-18 (Grism 8) from left to right. The solid black line is the object star and the solid red line is the reference star.
}\label{stars_spectra}
\end{figure*}

Initial data reduction was carried out using IRAF (Image Reduction and Analysis Facility), which includes bias correction, flat fielding, and optimal aperture extraction. The dispersion solution for the wavelength calibration is derived from the arc lamps (FeAr and FeNe for Grism 7 and Grism 8, respectively). The calibration lamp exposures were taken during the beginning and end of the transit observations with the same setup as the science exposures, but using a narrower slit of width 1.92\arcsec. Figure \ref{stars_ccd_image} shows the 2D image of the spectra for all three objects. Since the on-sky observations were taken continuously (without calibration lamp exposures) for several hours, a wavelength drift of the spectra is expected due to instrument flexure, temperature changes, and seeing. 
Any shifts between the first exposure (that is, taken closer to the lamp spectrum) and subsequent exposures are corrected by finding the offset and applied to the dispersion solution.
We used the centroid shift in the prominent telluric lines in each exposure to estimate the offset. Visually inspecting the individual spectra by overplotting with the first exposure seems to be well aligned within a pixel accuracy. The wavelength shift from first to last exposure for the three targets is less than 4 pixels or 3 {\AA}.
The wavelength offset between the reference and target star spectra on the detector was also adjusted for each exposure. The typical offset is 1.2 {\AA}, 2.15{\AA}, and 6.5{\AA} for WASP-127, HAT-P-1A, and KELT-18, respectively. Figure \ref{stars_spectra} represents the wavelength-calibrated spectra of the target star and reference from the same exposure.

\subsection{Constructing the white lightcurve} \label{cmc}
\begin{figure*}
\centering
\includegraphics[width=0.7\columnwidth]{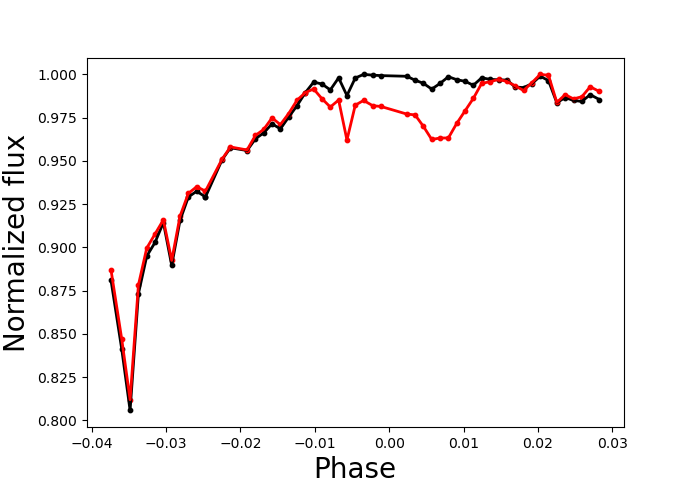}
\includegraphics[width=0.67\columnwidth]{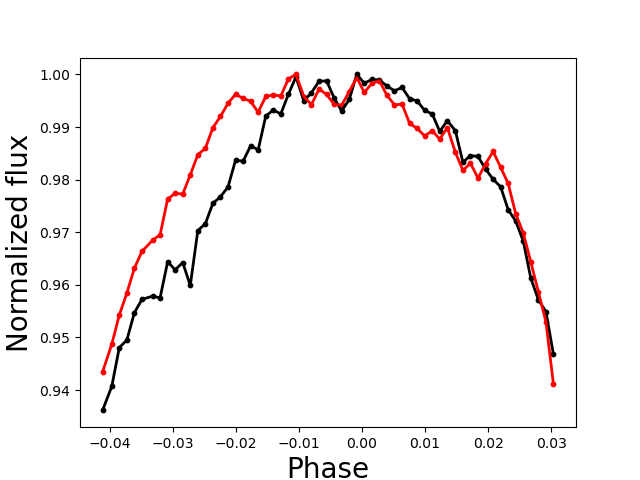}
\includegraphics[width=0.67\columnwidth]{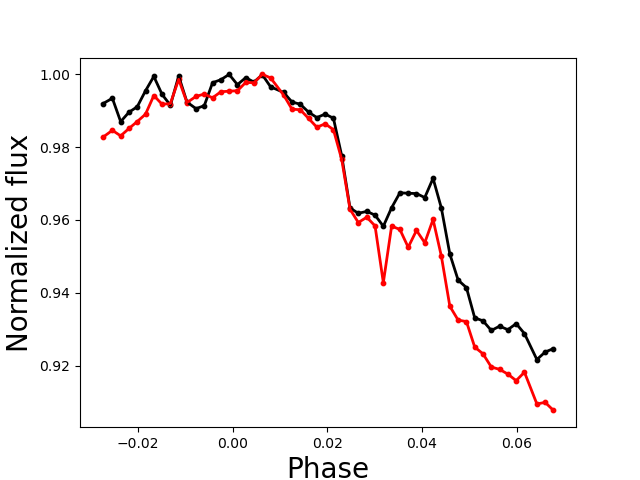}
\caption{Individual lightcurves of object star (black)  and the reference star (red) follows similar trend as a function of time. HAT-P-1A, WASP-127, and KELT-18 from left to right, respectively.}
\label{individual_lightcurve}
\end{figure*}

A transit white lightcurve was constructed by integrating the flux over the observed wavelength range of both the target and the reference star in each exposure. Individual white lightcurve of the target and the reference star are shown in Figure \ref{individual_lightcurve}. A differential white lightcurve is constructed by dividing the target star lightcurve by the reference star lightcurve. This technique assumes that the time-dependent systematic errors are the same in the observed flux of both the target and reference stars. We removed all the telluric regions when constructing the white lightcurve to avoid introducing additional systematic errors.

The best-fit model to the lightcurve was obtained using {\it PyTransit \footnote{\href{https://github.com/hpparvi/PyTransit}{https://github.com/hpparvi/PyTransit}}} \citep{Parviainen_2015}, a python-based lightcurve modeling tool widely used for exoplanet detection and characterization \citep{Jiang_2022, Casasayas_2022}. {\it PyTransit} provides efficient CPU and GPU implementations and combines a Bayesian approach to inference with Markov chain Monte Carlo (MCMC) \citep{Foreman-Mackey_2013} sampling for the posterior estimation. This package includes various transit models like \citet{Mandel_2002,Gimenez_2006,Maxted_2019,Parviainen_2020} with various limb darkening laws. Out of this, we used the popular Mandel-Agol analytic transit lightcurve model with quadratic limb darkening law. The primary input to {\it Pytransit} are the normalized, detrended flux in the user-defined wavelength region and the observation time of each exposure. A second-order polynomial function is used for detrending the lightcurve in time. The mid-transit time ($T_{c}$) and the transit depth from previous studies were used as prior parameters for the subsequent MCMC  run (30 chains of 2000 iterations) to obtain a reasonable posterior distribution of the parameters using the Bayesian approach. The adopted mid-transit time for HAT-P-1b, WASP-127b, and KELT-18b are $2453979.932$ \citep{Nikolov_2014}, $2458293.252$ \citep{Spake_2021} and $2457542.525$ \citep{McLeod_2017}, respectively. We used a transit depth of $0.01392$ \citep{Nikolov_2014} , $0.01013$ \citet{Spake_2021} and $0.00716$ \citep{McLeod_2017} for HAT-P-1b, WASP-127b, and KELT-18b, respectively. These initial values are optimized with other parameters like the orbital inclination, impact parameter, and the stellar density. The orbital period of all three targets are fixed in the analysis. We used an orbital period of $4.46529976$ days \citep{Nikolov_2014}, $4.17807015$ days \citep{skaf_2020}, and $2.8717510$ days \citep{McLeod_2017} for HAT-P-1b, WASP-127b and KELT-18b, respectively. The orbital eccentricity of all three targets was fixed at zero, assuming a circular orbit. The quadratic limb-darkening coefficients are estimated for each target using {\it Pytransit} and fixed for a model. Along with the fixed parameters, Table \ref{table_parameter} shows all the transit parameters obtained from the best-fit white lightcurves for all three targets. For example, the posterior density distribution of different parameters for HAT-P-1b is shown in Figure \ref{posterior_hatp1} in the Appendix. 

\begin{table*}
\begin{center}
\caption{The best-fit parameters from white lightcurve using the posterior density distribution. The orbital period (P) in days, the mid-transit time $T_{c}$ in JD, orbital inclination (i) in degrees, impact parameter (b), qudratic limb darkening coefficents u and v, orbital eccentricity (e), and the planet to star radius ratio ($R_{p}/{R_{*}}$) are listed below.}
\begin{tabular}{ccrrrrrrrrrrrr}
\hline\hline
Parameters&HAT-P-1b&WASP-127b&KELT-18b&\\
\hline
P (days) &4.4652&4.1780  &2.8716 & fixed\\
$T_{c}$ &2459079.3110$\pm$0.0003&2458552.3070 $\pm$ 0.0321&2458645.2389 $\pm$ 0.0014&fitted\\
i (degree)& 85.37 $\pm$ 0.8594 & 89.38 $\pm$ 0.3036 &89.15 $\pm$ 1.7188 & fitted\\
b & 9.85 $\pm$ 0.2391 & 7.89 $\pm$0.0101 &5.14 $\pm$ 0.0101 &fitted\\
u &0.3005 & 0.6550&0.2704 &fixed\\
v &0.4913& 0.5399&0.3904 &fixed\\
e&0&0&0&fixed\\
$R_{p}/{R_{*}}$ & $0.1122 \pm0.0015$ & $0.0964 \pm0.0016$ & $0.0902 \pm0.0017$ & fitted \\
\hline
\label{table_parameter}
\end{tabular}
\end{center}
\end{table*}

\subsection{Constructing low-resolution transmission spectra}
\begin{figure*}
\centering
\includegraphics[width=0.68\columnwidth]{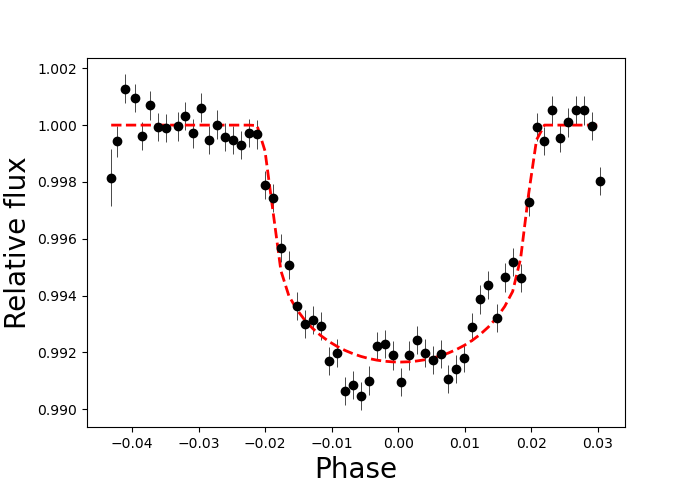}
\includegraphics[width=0.68\columnwidth]{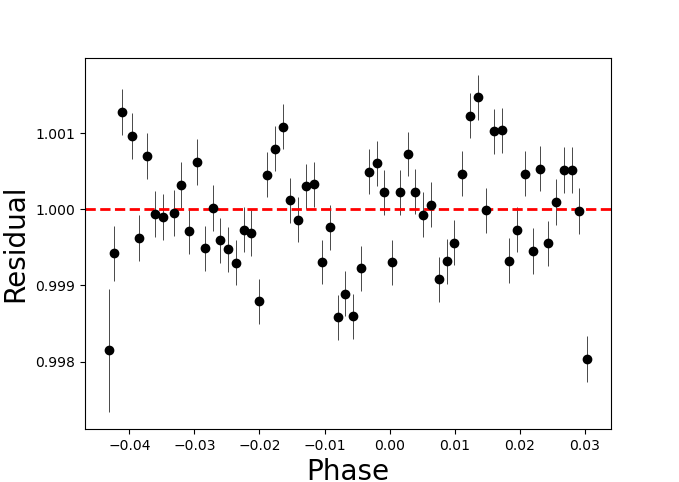}
\includegraphics[width=0.68\columnwidth]{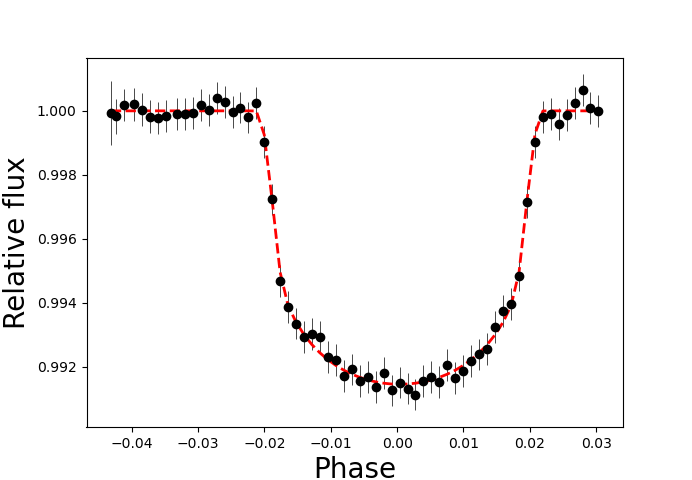}
\caption{Observed white lightcurve (scattered point) of WASP-127 with the best-fit model (red dashed line) in left after detrending in time. The residual plot in the center and the lightcurve after applying the common mode correction is on the right.}\label{demo_method_wasp127}
\end{figure*}

\begin{figure*}
\centering
\includegraphics[width=0.7\columnwidth]{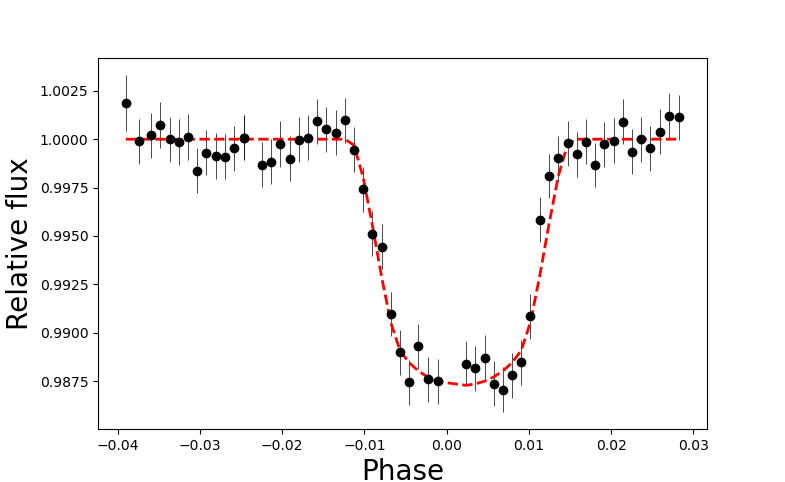}
\includegraphics[width=0.62\columnwidth]{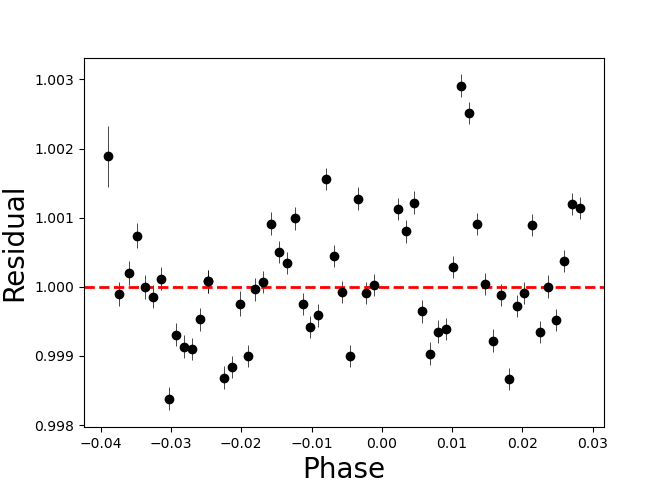}
\includegraphics[width=0.7\columnwidth]{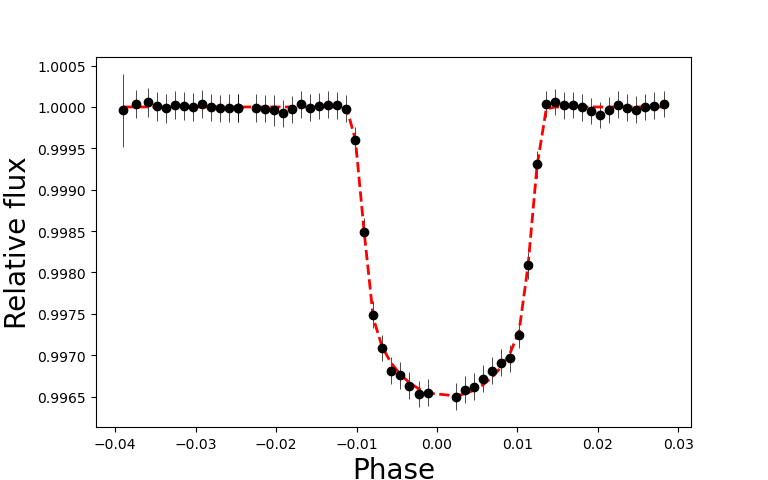}
\caption{Observed white lightcurve (scattered points) of HAT-P-1b with the best-fit model (red dashed line) in left after detrending in time. The residual plot in the center and the lightcurve after applying the common mode correction is on the right.}\label{demo_method_hatp1}
\end{figure*}

\begin{figure*}
\centering
\includegraphics[width=0.66\columnwidth]{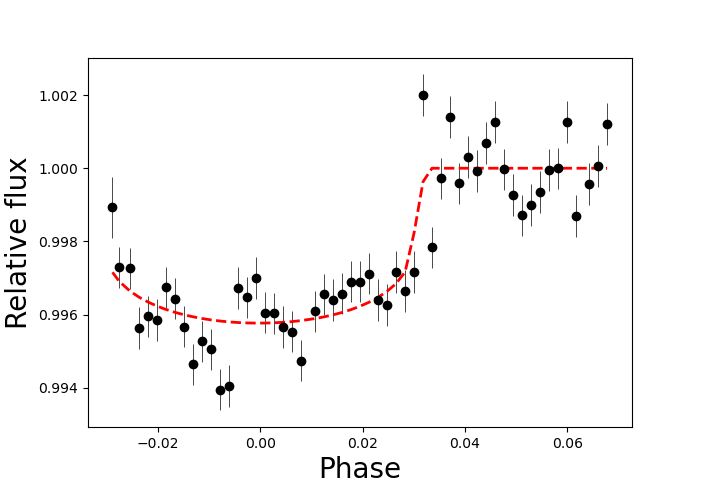}
\includegraphics[width=0.66\columnwidth]{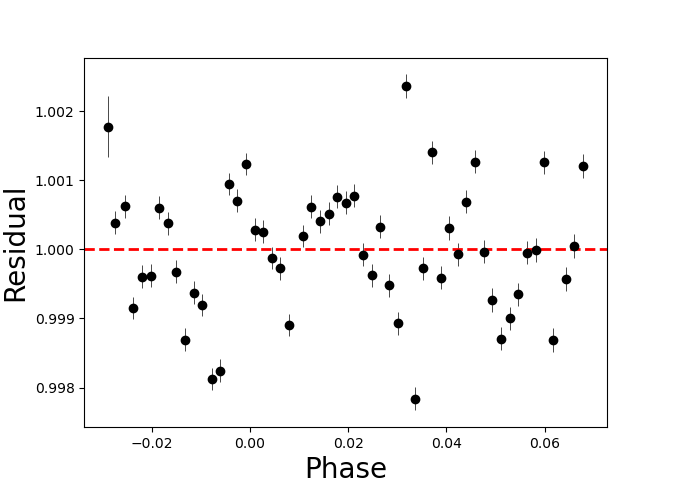}
\includegraphics[width=0.74\columnwidth]{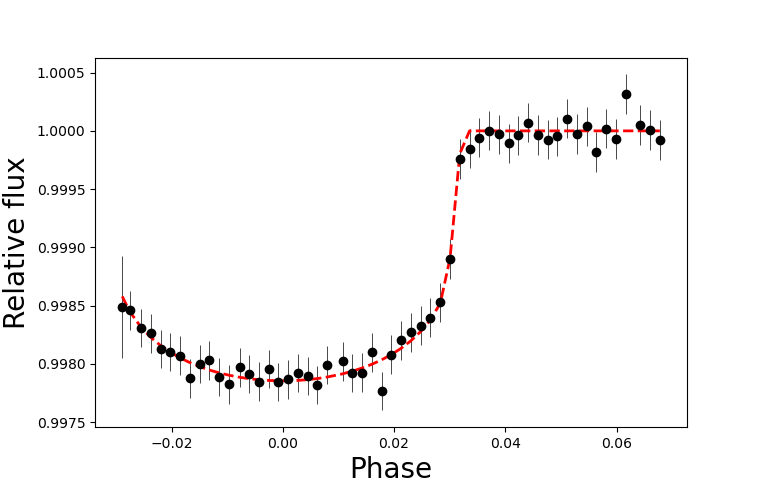}
\caption{Observed white lightcurve (scattered points) of KELT-18 with the best-fit model (red dashed line) in left after detrending in time. The residual plot in the center and the lightcurve after applying the common mode correction is on the right. The ingress part of the transit is missed due to instrument failure during the observation.}\label{demo_method_kelt18}
\end{figure*}

We applied the Common Mode Correction (CMC) technique \citep{Gibson_2013,Stevenson_2014,Weaver_2020,Wilson_2020} to remove dominant time-dependent systematics (common across all wavelengths) from the individual spectroscopic lightcurves. This entailed dividing the data into wavelength bins such that the SNR of individual spectroscopic lightcurves was always 1000. Then, we calculated the residuals from the white lightcurve fits (discussed in Section \ref{cmc}) - that is, the observed white lightcurve divided by the best-fit model. This is the “common mode” noise. We  divided each of the spectroscopic lightcurves by the common mode noise to construct the common mode corrected lightcurves. By employing CMC, we sacrifice the information of the absolute transit depths and instead obtain relative transit depths, but this approach will avoid most of the time dependent scatter in the lightcurves and improve the precision of the measured transit depths in all wavelength channels. We fit the individual wavelength-dependent lightcurves using {\it PyTransit} following the similar procedure explained in Section \ref{cmc} with the parameters from the best-fit white lightcurve as prior input. Figure \ref{demo_method_wasp127}, Figure \ref{demo_method_hatp1}, and Figure \ref{demo_method_kelt18} demonstrate the method for constructing the common mode corrected spectroscopic lightcurves for WASP-127b, HAT-P-1b, and KELT-18b, respectively (a typical single wavelength bin of SNR=1000 is shown for each). 

\begin{figure*}
\centering
\includegraphics[scale=0.65]{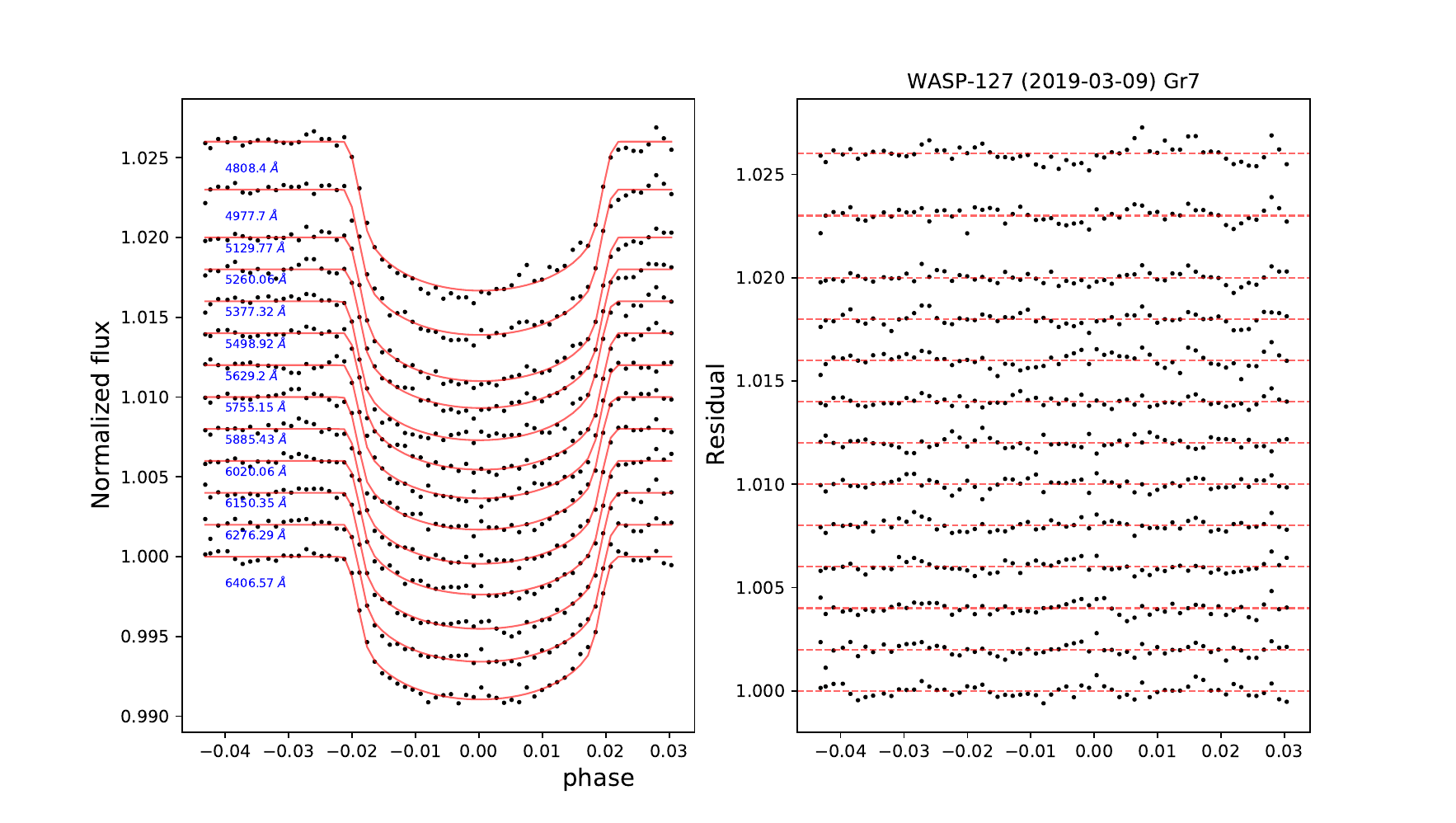} 
\caption{ Grism 7 lightcurves in each wavelength bin (scattered points) and the best-fit transit models (solid lines) stacked from top to bottom for WASP-127b. A constant offset of $0.01$  is added to the lightcurves in y-direction  for clear separation. A wavelength bin of constant SNR=1000 is used for constructing the lightcurves.}
\label{wasp_kelt_lightcurve}
\end{figure*}

\begin{figure*}
\centering
\includegraphics[scale=0.6]{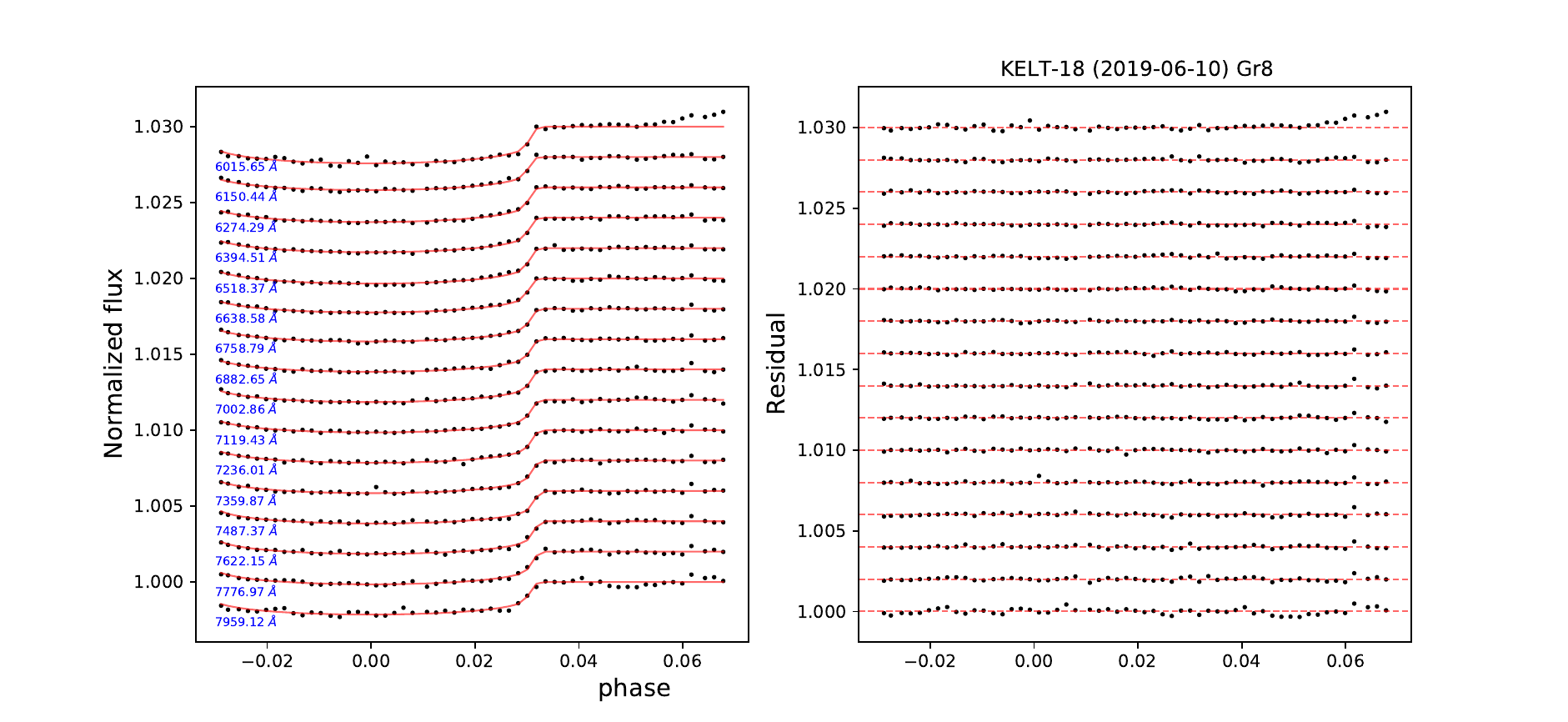}
\caption{Grism 8 lightcurves in each wavelength bin (scattered points) and the best-fit transit models (solid lines) stacked from top to bottom for KELT-18b. A constant offset of $0.01$  is added to the lightcurves in y-direction  for clear separation. A wavelength bin of constant SNR=1000 for constructing the lightcurves.}
\label{kelt18_kelt_lightcurve}
\end{figure*}

\begin{figure*}
\centering
\includegraphics[scale=0.6]{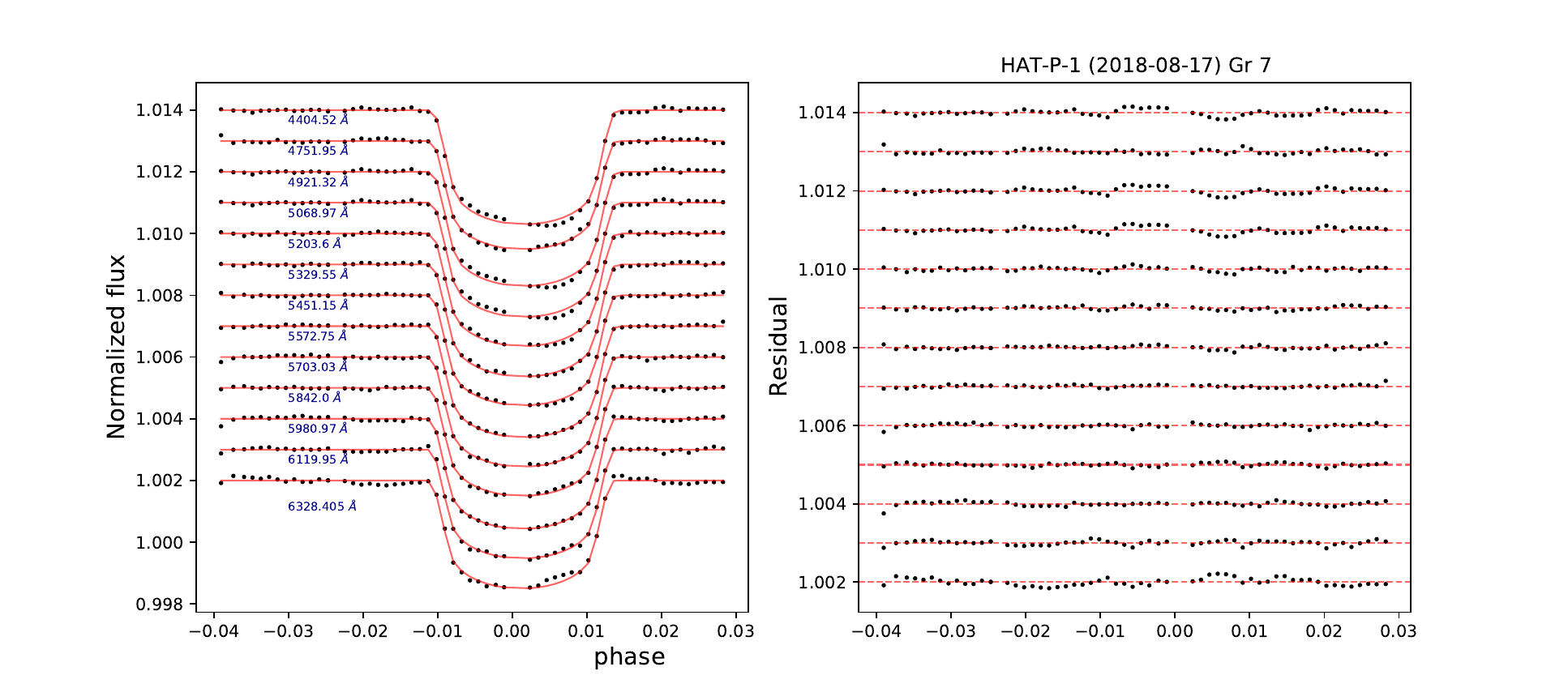} 
\caption{Grism 7 lightcurves in each wavelength bin (scattered points) and the best-fit transit models (solid lines) stacked from top to bottom for HAT-P-1b. 
A constant offset of $0.01$  is added to the lightcurves in y-direction  for clear separation. A wavelength bin of constant SNR=1000 is used for constructing the lightcurves.}
\label{hatp1_lightcurve}
\end{figure*}

Chromatic dispersion and achromatic differential refraction are the two different components that contribute to slit loss during the observation. As the chromatic dispersion is constant for a field, a differential analysis can minimize the effects even though the ideal situation is to have an Atmospheric Dispersion Corrector (ADC). All three objects and the corresponding reference stars have similar physical parameters (see Table \ref{table2}), so it is unlikely that the differential atmospheric dispersion will contribute to these observations. The field differential refraction is due to the airmass variation in the Field Of View (FOV). We estimated the apparent shift in the position of the stars due to differential refraction using Equation 3 in \citet{Janssen_2014} for the airmass ranges in our study (Table \ref{table1}) and considering the binary separation (Table \ref{table2}). It is $0.005\arcsec-0.002\arcsec$ for HAT-P-1A, $0.03\arcsec-0.03\arcsec$ for WASP-127, and $0.09\arcsec-0.2\arcsec$ for KELT-18 at the beginning and end of the observation. The values are smaller than the slit width ($15.4 \arcsec$), so differential refraction is unlikely to play a significant role in these observations. Any significant variations as a function of time would be due to the instrumental effects, which can be removed using CMC. The common mode corrected lightcurve at different wavelength bins for all three targets presented in Figure \ref{wasp_kelt_lightcurve}, Figure \ref{kelt18_kelt_lightcurve}, and Figure \ref{hatp1_lightcurve}. The transit depth measured at different wavelength bins is presented in Table \ref{table_kelt18}, Table \ref{table_hatp1}, and Table \ref{table_wasp127} for KELT-18b, HAT-P-1b, and WASP-127b, respectively. 

\begin{table}
\begin{center}
\caption{Observed transit depth of constant SNR=1000 for KELT-18b. The central wavelength (W) in \AA, width of each wavelength bin (\AA), the planet to star radius ratio ($R_{p}/{R_{*}}$), and error in the estimate $R_{p}/{R_{*}}$ are presented in columns one to four, respectively. }
\begin{tabular}{crrrrrrrrrrrr}
\hline\hline
W (center)&W (width)&$R_{p}/{R_{*}}$ &$R_{p}/{R_{*}}$(err)\\
\hline
7959.12&	100.18&	0.07347&	0.00255 \\
7776.97&	81.96&	0.07343&	0.00242 \\
7622.15&	72.85&	0.07452&	0.00229 \\
7487.37&	61.93&	0.07211&	0.00126 \\
7359.87&	61.93&	0.07215&	0.00188 \\
7236.01&	58.29&	0.07146&	0.00148 \\ 
7119.43&	58.43&	0.07229&	0.00162 \\
7002.86&	58.28&	0.07179&	0.00158 \\
6882.65&	61.93&	0.07164&	0.00121   \\ 
6758.79&	61.92&	0.07158&	0.00115 \\
6638.58&	58.28&	0.07252&	0.00110 \\
6518.37&	61.93&	0.07396&	0.00120 \\
6394.51&	61.93&	0.07343&	0.00115 \\
6274.29&	58.28&	0.07325&	0.00102 \\
6150.44&	66.37&	0.07142&	0.00098 \\
\hline
\label{table_kelt18}
\end{tabular}
\end{center}
\end{table}

\begin{table}
\begin{center}
\caption{Observed transit depth of constant SNR=1000 for HAT-P-1b. The central wavelength (W) in \AA, width of each wavelength bin (\AA), the planet to star radius ratio $R_{p}/{R_{*}}$, and error in the estimate ($R_{p}/{R_{*}}$) are presented in columns one to four, respectively. }
\begin{tabular}{crrrrrrrrrrrr}
\hline\hline
W (center)&W (width)&$R_{p}/{R_{*}}$ &$R_{p}/{R_{*}}$(err)\\
\hline
6328.405&	138.8&	0.11998&	0.00197 \\
6119.95&	69.4&	0.11961&	0.00198 \\
5980.97&	69.4&	0.12031&	0.00188 \\
5842.0&	    69.4&	0.11932&	0.00191 \\
5703.03&	69.4&	0.12008&	0.00186 \\
5572.75&	60.8&	0.12119&	0.00202 \\
5451.15&	60.8&	0.12118&	0.00199 \\
5329.55&	60.8&	0.12212&	0.00208 \\
5203.6&	    65.1&	0.12295&	0.00218 \\
5068.97&	69.4&	0.12410&	0.00224 \\
4921.32&	78.1&	0.12079&	0.00220 \\
4751.95&	91.2&	0.12563&	0.00223 \\
4404.52&	256.2&	0.12134&	0.00296 \\
\hline
\label{table_hatp1}
\end{tabular}
\end{center}
\end{table}

\begin{table}
\begin{center}
\caption{Observed transit depth of constant SNR=1000  for WASP-127b. The central wavelength (W) in \AA, width of each wavelength bin (\AA), the planet to star radius ratio ($R_{p}/{R_{*}}$), and error in the estimate $R_{p}/{R_{*}}$ are presented in columns one to four, respectively.}
\begin{tabular}{crrrrrrrrrrrr}
\hline\hline
W (center)&W (width)&$R_{p}/{R_{*}}$ &$R_{p}/{R_{*}}$(err)\\
\hline
6406.57&	69.48&	0.09747&	0.00117 \\
6276.29&	60.8&	0.09702&	0.00136  \\
6150.35&	65.14&	0.09676&	0.00115 \\
6020.06&	65.14&	0.09618&	0.00140 \\
5885.43&	69.48&	0.09661&	0.00133 \\
5755.15&	60.8&	0.09619&	0.00083 \\
5629.2&	    65.1&	0.09615&	0.00185 \\
5498.92&	65.1&	0.09681&	0.00117 \\
5377.32&	56.4&	0.09758&	0.00138 \\
5260.06&	60.8&	0.09728&	0.00167 \\
5129.77&	69.4&	0.09838&	0.00160 \\
4977.7&	    82.5&	0.09892&	0.00243 \\
4808.4&	    86.8&	0.09955&	0.00210 \\
\hline
\label{table_wasp127}
\end{tabular}
\end{center}
\end{table}

\section{Modelling the transmission spectra} \label{model}
\begin{figure}
\centering
\includegraphics[scale=0.5]{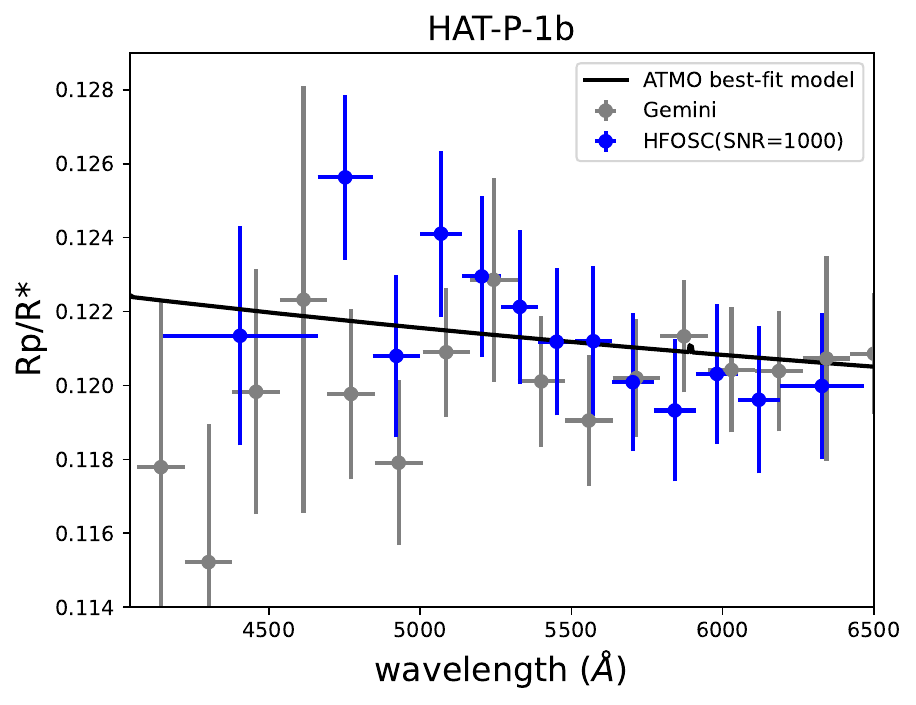} 
\caption{Figure showing best-fit ATMO model (black) with HCT/HFOSC transmission spectra observations of HAT-P-1b (blue). The earlier observation from 8.1m Gemini/GMOS is plotted in grey \citep{Todorov_2019}.}
\label{hatp1_trans1_model}
\end{figure}

\begin{figure}
\centering
\includegraphics[scale=0.5]{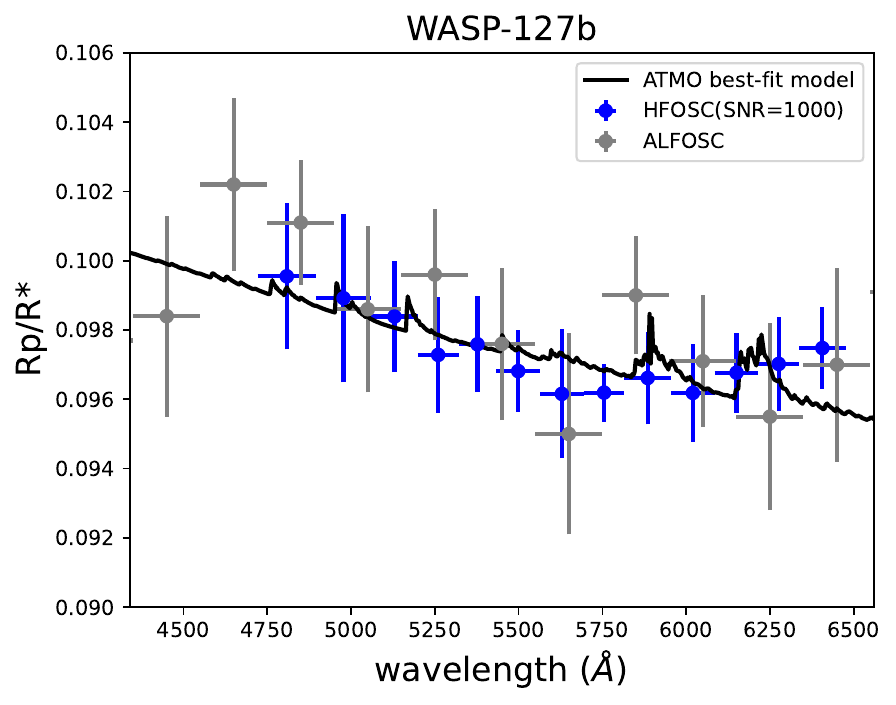} 
\caption{Figure showing best-fit ATMO model (black) with HCT/HFOSC transmission spectra observations of WASP-127b (blue). The earlier observation from 2.5 m NOT/ALFOSC is plotted in grey \citep{Palle_2017}.}
\label{wasp127_trans1_model}
\end{figure}

To interpret the observed transmission spectra of HAT-P-1b and WASP-127b, we used a grid of chemical-equilibrium model atmospheres and their calculated transmission spectra, which were developed using ATMO \citep{Goyal_2017}. ATMO is a 1D planetary atmosphere model that can compute radiative-convective equilibrium pressure-temperature ($P$-$T$) profiles consistent with equilibrium as well as dis-equilibrium chemistry \citep{Amundsen2014,Tremblin2015,Drummond2016,Goyal_2017, Goyal_2020}. The transmission spectra model grid used in this work assumes an isothermal $P$-$T$ profile and equilibrium chemistry with condensation as described in detail in \citet{Goyal_2017}. The grid consists of a total of 3920 model transmission spectra for each planet for a range of isothermal temperatures (T$_{\rm eq}$ - 300 to T$_{\rm eq}$ + 300), metallicity (0.005-200x solar), C/O ratio (0.15-1.5). Haze is parameterized using a multiplicative haze factor ranging from 1 to 1100 times the Rayleigh scattering opacity due to gases in the atmosphere. Clouds are treated as large particles and parameterized with an additive grey opacity ranging from 0 (no cloud) to 1 times the H$_2$ Rayleigh scattering cross section at 350 nm. A detailed explanation of the model and grid parameters is provided in \citet{Goyal_2017}. The spectral resolution of the model used for this analysis is  R=5000 at 0.2 $\mu$m while decreasing to R=100 at 10 $\mu$m. The best-fit models are determined by $\chi^2$ minimization by re-binning the model spectra to the observed data. This model grid has been extensively used to interpret several ground-based observations \citep[For e.g.][]{Carter_2020, Wilson_2020,Spyratos_2021,Weaver_2021}. The Figure \ref{hatp1_trans1_model} and Figure \ref{wasp127_trans1_model} show the observed transmission spectra with the corresponding best-fit model spectra for HAT-P-1b and WASP-127b, respectively. 

\begin{figure}
\centering
\includegraphics[scale=0.55]{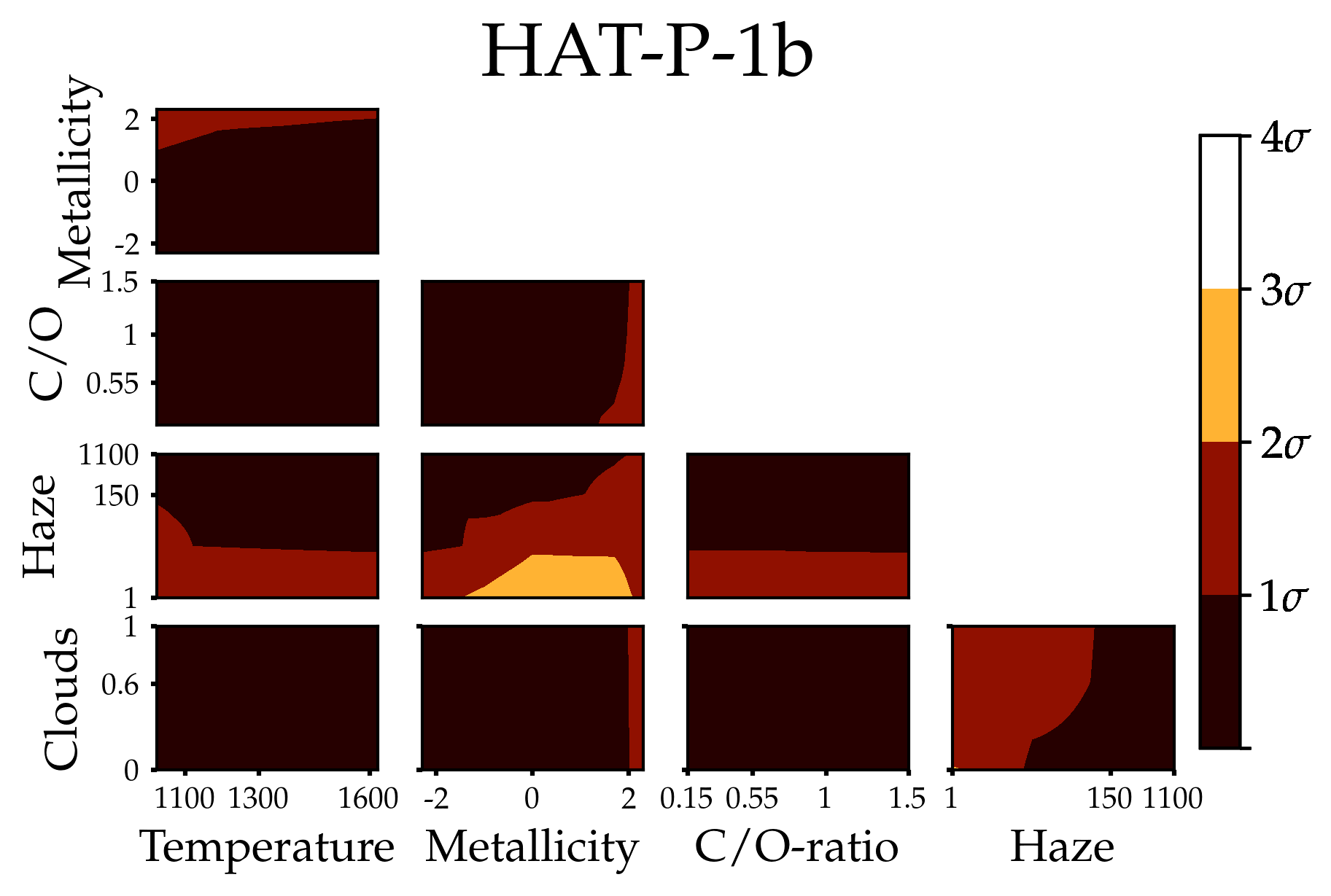} 
\caption{Figure showing $\chi^2$ map for HAT-P-1b. Contours of $\chi^2$ are shown for all the combinations of grid parameters. Cloud and haze factor axes are log-scaled. Metallicity is also log-scaled, 0 being solar metallicity and 2 being 100 times solar metallicity. The colors indicate confidence intervals as shown in colormap to the right.}
\label{hatp1b_chimap_model}
\end{figure}

For HAT-P-1b, we find that the best-fit model has a temperature of 1622\, K, 0.005x solar metallicity, C/O ratio of 1.5, Haze-factor of 1100, and cloud-factor of 0. The best-fit model has $\chi^2$ = 6.67 with 0.75 $\leq$ P($\chi^2$) $\leq$ 0.9. We could not find evidence of any molecular/atomic species with the current dataset. We show the 3$\sigma$ range of the model parameters with respect to the best-fit model with a $\chi^2$ map for HAT-P-1b in Figure \ref{hatp1b_chimap_model}. The methodology for plotting this $\chi^2$ map is the same as discussed in \citet{Goyal_2017}. From these plots, we can infer that we are not able to constrain metallicity, C/O ratio, and clouds robustly for HAT-P-1b with the HFOSC observations. The only physical characteristic that we were able to constrain to some extent is the Rayleigh Scattering slope, with the presence of enhanced Rayleigh Scattering haze, which for the best-fit model is 1100 times the nominal Rayleigh scattering, with values between 10x to 1100x (logarithmically spaced) within the 1$\sigma$ range of the minimum $\chi^2$ model. 

\begin{figure}
\centering
\includegraphics[scale=0.6]{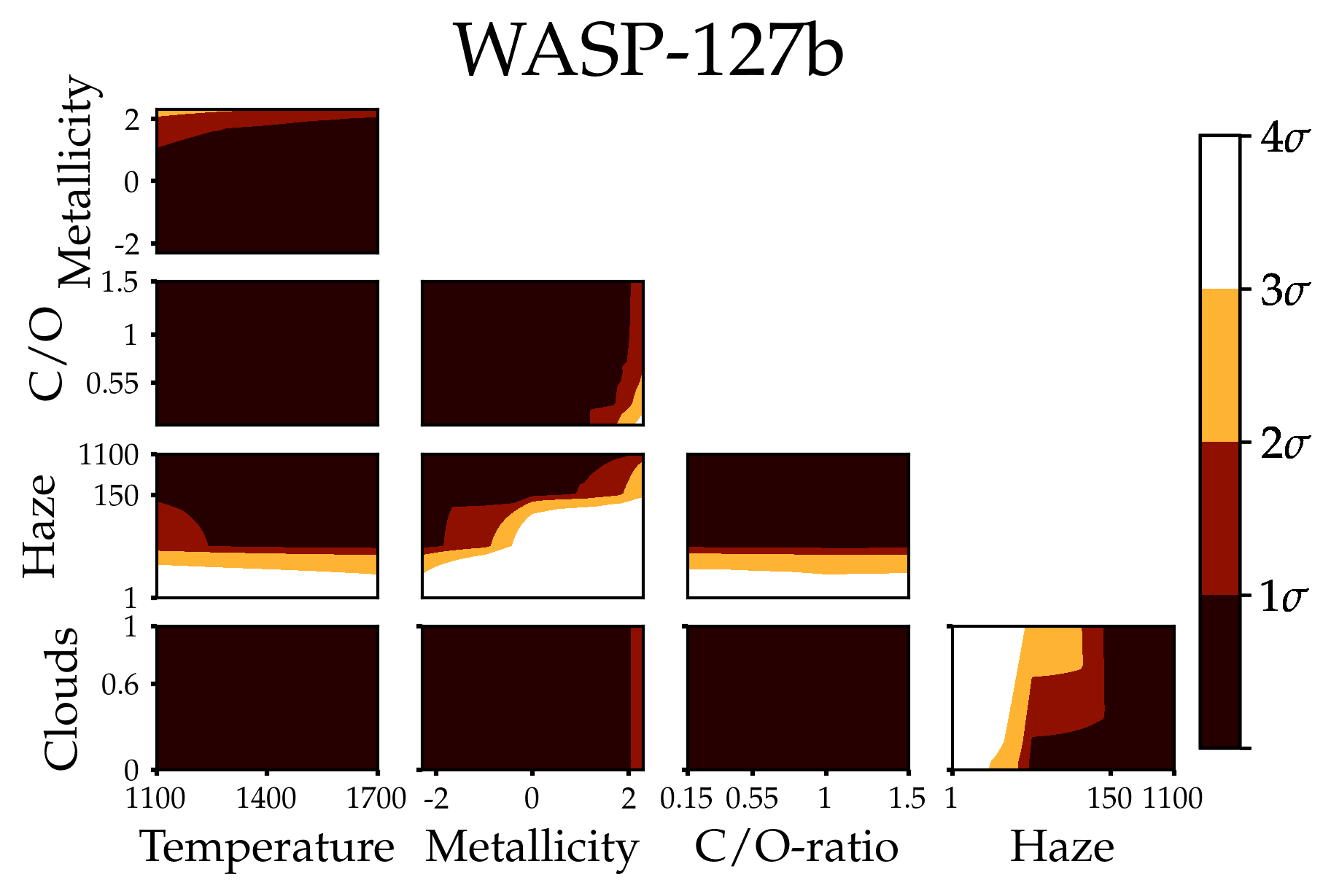} 
\caption{Figure showing $\chi^2$ map for WASP-127b. Contours of $\chi^2$ are shown for all the combinations of grid parameters. Cloud and haze factor axes are log-scaled. Metallicity is also log-scaled, 0 being solar metallicity and 2 being 100 times solar metallicity. The colors indicate confidence intervals as shown in colormap to the right.}
\label{wasp127_chimap_model}
\end{figure}

For WASP-127b, we find that the best-fit model has a temperature of 1700\, K, 0.005x solar metallicity, C/O ratio of 0.35, Haze-factor of 150, and cloud-factor of 1. The best-fit model has $\chi^2$ = 5.71 with 0.95 $\leq$ P($\chi^2$) $\leq$ 0.99. For WASP-127b, too, we are only able to constrain the Rayleigh scattering slope,  with the presence of enhanced Rayleigh scattering haze. The haze factor for the best-fit model is 150 times the nominal Rayleigh scattering, with values between 10x to 1100x within the 3$\sigma$ range of the minimum $\chi^2$ model. However, as metallicity increases, the haze factor is strongly constrained to values larger than 150x. For metallicities greater than 100x solar, the haze factor lies between 150x to 1100x, that too within the 3$\sigma$ range of the minimum $\chi^2$ model as shown in the $\chi^2$ map for WASP-127b (Figure \ref{wasp127_chimap_model}). The large value of the haze factor implies a substantially hazy atmosphere, which is consistent with interpretation using GTC observations from \citet{Chen_2018} but inconsistent with HST observations from \citet{Spake_2021}. However, we do not detect the presence of Na as in \citet{Chen_2018} and \citet{Spake_2021} due to the lower precision and spectral resolution of the HFOSC data. The best-fit model predicts TiO/VO spectral features similar to \citet{Palle_2017}, but the precision and wavelength coverage of the current dataset do not allow us to constrain it in any way. 

\begin{figure}
\centering
\includegraphics[scale=0.5]{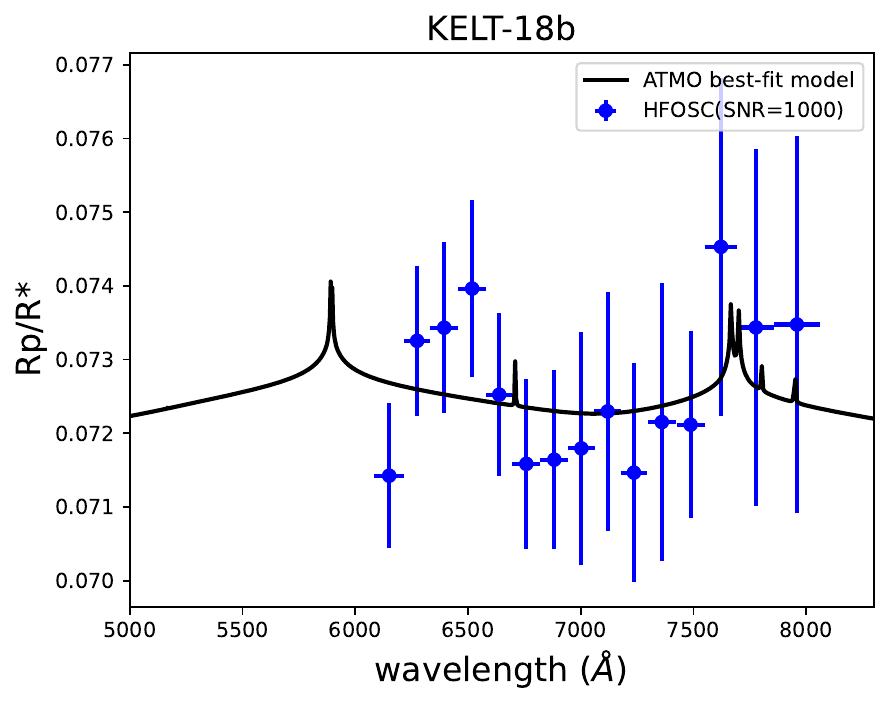} 
\caption{Best-fit ATMO model (black) with HCT/HFOSC transmission spectra observations of KELT-18b (blue).}
\label{kelt18_trans1_model}
\end{figure}

Since for KELT-18b, the model transmission spectra were not available in the planet-specific library presented in \citet{Goyal_2017}, we, therefore, used ATMO generic grid scalable model spectra as presented in \citet{Goyal_2019a}. This model grid can be used to scale to any planet's transmission spectra with H$_2$-He dominated atmosphere across a range of temperatures (400 - 2600\, K), planetary gravities (5 - 50 m/s$^2$), atmospheric metallicities (1x - 200x solar), C/O ratio (0.35 - 1), haze-factor (1-1100) and cloud-factor (0 - 1). This model grid also assumes isothermal $P$-$T$ profiles. For KELT-18b, the best-fit model has a temperature of 1200\, K, 10x solar metallicity, C/O ratio of 1 without any haze and clouds. The best-fit model has $\chi^2$ = 6.32 with 0.90 $\leq$ P($\chi^2$) $\leq$ 0.95. Due to the extremely low precision of this KELT-18b data, we are not able to constrain any physical or chemical characteristics with our models. The best-fit ATMO model, along with the observed HFOSC data for KELT-18b, is shown in Figure \ref{kelt18_trans1_model}.

To demonstrate the advantage of HFOSC observations and place more robust constraints on the estimated parameters for both HAT-P-1b and WASP-127b, we fitted the model grid with the HFOSC observations along with available high quality HST/Spitzer observations in the infrared. Figure \ref{hatp1_hfosc_hst_model} in Appendix shows the best-fit model from the HAT-P-1b model grid, when fitted to HFOSC observations combined with HST observations from \citet{Wakeford_2013}. The offset between HFOSC and HST is a free parameter when fitting models to observations. The best-fit model gives an equilibrium temperature of 1472\, K,  solar metallicity, C/O ratio of 0.15 (sub-solar), Haze-factor of 1100, and cloud-factor of 0 (no clouds). The best-fit model has a $\chi^2$ value of 32.57 with 39 degrees of freedom. The best-fit offset value between HFOSC and HST observations is -0.0029. Figure \ref{hatp1_corner_hst_model}  and \ref{hatp1_corner_hfosc_hst_model} in the Appendix show the $\chi^2$ maps for HAT-P-1b model grid when fitted to just HST infrared observations and HFOSC plus HST observations, respectively. It can be clearly seen from the $\chi^2$ maps that the 3 $\sigma$ constraints on the model parameters are better with HFOSC plus HST infrared observations as compared to just HST infrared observations and HFOSC observation. Specifically, we are able to put robust constraints on the haze parameter, as expected, since we can constrain the Rayleigh scattering slope of the transmission spectrum with HFOSC observations. 
The combined spectrum constrains the C/O ratio to be less than 0.7 and the metallicity to be within 0.1x and 50x solar metallicity. It also helps place stronger  constraints on the equilibrium temperature of the planet, with preference towards higher equilibrium temperature, as seen in the $\chi^2$ maps. While HST infrared observations help in placing stronger constraints on the C/O ratio and metallicity, HFOSC observations help in placing stronger constraints on the haze parameter and temperature. 

Figure \ref{wasp127_hfosc_hst_model} in the Appendix shows the best-fit model from the WASP-127b model grid when fitted to HFOSC observations combined with infrared observations from HST and SPITZER \citep{Spake_2021}. We find that the best-fit model has a temperature of 1100\, K,  10$\times$ solar metallicity, solar C/O ratio, Haze-factor of 150, and a cloud-factor of 1.0. The best-fit model has a $\chi^2$ value of 57.16 with 42 degrees of freedom. The best-fit offset value between HFOSC and HST/SPITZER observations is 0.0048. Figure \ref{wasp127_corner_hst_model}  and \ref{wasp127_corner_hfosc_hst_model} in the Appendix show the $\chi^2$ maps for WASP-127b model grid when fitted to just infrared HST/SPITZER observations and HFOSC plus HST/SPITZER observations, respectively. It can be clearly seen from the $\chi^2$ maps that the 3 $\sigma$ constraints on the model parameters are better with HFOSC plus HST/SPITZER observations as compared to just HST/SPITZER observations in the infrared. Similar to HAT-P-1b, the constraints are worse with just HFSOC observations, significantly better with just HST/SPITZER observations and best with the combined observed spectra. With just HFOSC observations we are able to constrain the haze parameter to be greater than 10x the multi-gas Rayeligh scattering while all the other model parameters are un-constrained. However, with just HST/SPITZER observations we obtain very strong 3$\sigma$ constraints on the equilibrium temperature (T$_{\rm eq}$ $\leq$ 1250), metallicity (1 $\leq$ Z $\leq$ 1.7), C/O ratio (0.35 $\leq$ C/O $\leq$ 0.56) and haze-factor (150 $\leq$ haze-factor $\leq$ 1100), without any constraints on the cloudiness factor. With the addition of HFOSC observations to this HST/SPITZER spectrum in IR, the constraints on the equilibrium temperature become more robust (T$_{\rm eq}$ $\leq$ 1100), the metallicity constraints remain same but the favoured metallicity (within 1$\sigma$) shifts to 10x solar, the haze factor constraints also remain same but the favoured haze-factor (within 1$\sigma$) shifts to 150x multi-gas Rayeligh scattering compared to 1100x with just HST/SPITZER observations. Additionally, with the combined observations, we are able to place constraints even on the cloud factor (cloud-factor $\geq$ 0.2), basically discarding the no-cloud scenario. In summary, although we are able to place robust constraints with just HST/SPITZER observations in IR as compared to just HFOSC observations for WASP-127b, these constraints become more robust when both spectra are combined, especially for estimating equilibrium temperature, metallicity, haze and clouds. 

\section{Discussion and conclusion} \label{discussion}
In this paper, we present the low-resolution transit spectroscopic observation from HCT/HFOSC instrument for the first time. The wavelength-dependent lightcurves (Figure \ref{wasp_kelt_lightcurve}, Figure \ref{kelt18_kelt_lightcurve}, and Figure \ref{hatp1_lightcurve} ) are relatively free from systematic errors after CMC  with the RMS amplitude of residuals only 0.002 times above the expected photon noise after detrending in time. Relatively good precision is achieved mainly because of the availability of a suitable reference star for all three targets otherwise the CMC can underestimate the error in the transit depth and can misinterpret the transmission spectra \citep{Jiang_2022}. The ideal reference star is a binary twin companion as in the case of HAT-P-1A and WASP-127. 

We have tried different SNR (SNR=200, SNR=500, and SNR=1000) bins to understand the optimal SNR to construct the transmission spectra. Although the number of data points in the spectra increases with lower SNR bins, the estimated error in the transit depth also increases. So, there should always be a balance between the number of data points required and the accuracy of the estimated transit depth needed. We explored the sodium doublet region separately for WASP-127b and HAT-P-1b, choosing a narrow bandwidth of constant SNR=500 to detect the possible detection of sodium around 5890 {\AA} and the possible detection of potassium around 7670 {\AA} region for KELT-18b following the method explained in Section \ref{method}. Unfortunately, we could not detect sodium or potassium above the uncertainty level. Therefore, we recommend using wavelength bins of 70{\AA} or SNR=1000 for studying systems similar to WASP-127b and HAT-P-1b with HCT/HFOSC. For example, Figure \ref{wasp127_3bin} in the Appendix presents the observed spectra of WASP-127b in all three bins (SNR = 200, SNR = 500 and SNR = 1000) using HFOSC.

Even though the HCT/HFOSC observations are in decent agreement with the previous ground-based observations with the same technique, we could not detect any atomic/molecular species using the best-fit ATMO models for all three targets. In the case of WASP-127 b, the equilibrium temperature of the planet from our best fit model is 1100 K, comparable to the equilibrium temperature (1400 K) derived by \citet{Palle_2017}. At such temperatures, the abundance of TiO and VO is expected to be very low from equilibrium chemistry calculations, and thus may not be detectable. The enhanced Rayleigh scattering observed in WASP-127b (Figure \ref{wasp127_trans1_model}) is qualitatively comparable to the results of \citet{Palle_2017, Chen_2018}. However, quantitatively the best-fit model for \citet{Palle_2017} has a Rayleigh enhancement factor of 5 while in \citet{Chen_2018} this value is in the range of $\sim$ 8500 - 250000. Quantitatively, the results differ due to differences in the adopted models and model choices. While \citet{Palle_2017} used forward models with and without TiO/VO with a maximum Rayleigh scattering enhancement of 5 times the nominal Rayleigh scattering, \citet{Chen_2018} used free retrievals with priors on the Rayleigh scattering enhancement factor ranging from 10$^{-4}$ to 10$^{10}$. In this work, we compute a forward model grid with equilibrium chemistry and Rayleigh scattering enhancement factor going from 1 to 1100 times the multi-gas Rayleigh scattering. 

The enhanced Rayleigh scattering slope has been observed for other planets like HD 189733b (Sing et al. 2011), WASP-94b \citep{Ahrer_2022}, WASP-127b \citep{Palle_2017}, WASP-21b \citep{Alderson_2020}, etc. This can be due to an additional opacity source like a mineral cloud \citep{rayleigh_model} or a photochemical hazes \citep{Kawashima_2019,Ohno_2020} or a combination of clouds and hazes \citep{Pont_2013}. Unocculted starspots can also mimic the Rayleigh scattering slope \citep{McCullough_2014,Rackham_2017}. However, \citet{Spake_2021} found that WASP-127 is a photometrically quiet star from the TESS lightcurves, thus implying that the stellar inhomogeneities are unlikely to be the case here.

Even though HFOSC observations of HAT-P-1b are in good agreement with Gemini/GMOS observations \citep{Todorov_2019} (Figure \ref{hatp1_trans1_model}), the Gemini/GMOS observations suffers from strong systematic effects leading to large uncertainties comparable to HCT/HFOSC. The observed equilibrium temperature from both HFOSC (1472 K) and Gemini/GMOS (1500 K) are also comparable to each other. We estimated the Transmission Spectrum Metric (TSM) values from \citet{Kempton_2018} for all three planets. The TSM values are proportional to the SNR expected from the transmission spectra considering the transit depth and the scale height excluding the potential impact of clouds. The TSM values for KELT-18b, WASP-127b, and HAT-P-1b are 77.30, 447.68, and 196.67, respectively. The lower TSM value of KELT-18b implies that the observed featureless spectra is a combination of poor signal-to-noise ratio and the inherently small-scale height of the planet even though the clouds or hazes can mute the absorption features. The lower SNR can be due to a less suitable comparison star and a larger separation between the star and the reference star. The equilibrium temperature determined by \cite{McLeod_2017} for KELT-18b is 2085 K, which is 885 K higher than the best-fit equilibrium temperature from our study.

Considering a visual magnitude of less than 12 and planets orbiting around main sequence dwarf stars with TSM greater than 99 and a transit duration under seven hours, and most importantly, ensuring a reference star with a maximum separation of 11\arcmin, we expect to characterize 25 more exoplanet atmosphere from HCT/HFOSC. In a broader sense, this proof-of-concept study paves the way for examining the atmospheres of exoplanets using other relatively small telescopes that are outfitted with comparable spectrographic instruments. Our analysis combining HFOSC observations with infrared observations from HST or/and SPITZER proves that HFOSC can be an alternative optical instrument to use in conjunction with space-based infrared instruments like HST, SPITZER and JWST.


\section*{Acknowledgements}
We thank the staff of IAO, Hanle, and CREST, Hosakote, that made these observations possible. The facilities at IAO and CREST are operated by the Indian Institute of Astrophysics, Bangalore. This analysis has made use of the following software and packages: (i) {\it Image Reduction and Analysis Facility} ({\it IRAF}; \cite{tody1993}; (ii) {\it PyTransit} \citep{Parviainen_2015} (iii) {\it NumPy } \citep{numpy}; (iv) {\it Matplotlib} \citep{matplotlib}; (v) {\it SciPy} \citep{Virtanen2020}; (vi) {\it Astropy} \citep{astropy}; (vii) {\it emcee} \citep{Foreman-Mackey_2013}. JG acknowledges support from the SERB SRG Grant SRG/2022/000727-G for this work. We also thank the anonymous referee for the valuable comments, which greatly improved the paper. 

\section*{Data Availability}
Spectroscopic data of all three targets will be provided by the corresponding author upon reasonable request.
\bibliographystyle{mnras}
\bibliography{ref} 

\appendix \label{appendix}
\renewcommand\thefigure{A\arabic{figure}} 
\begin{figure}
\centering
\includegraphics[width=0.5\textwidth]{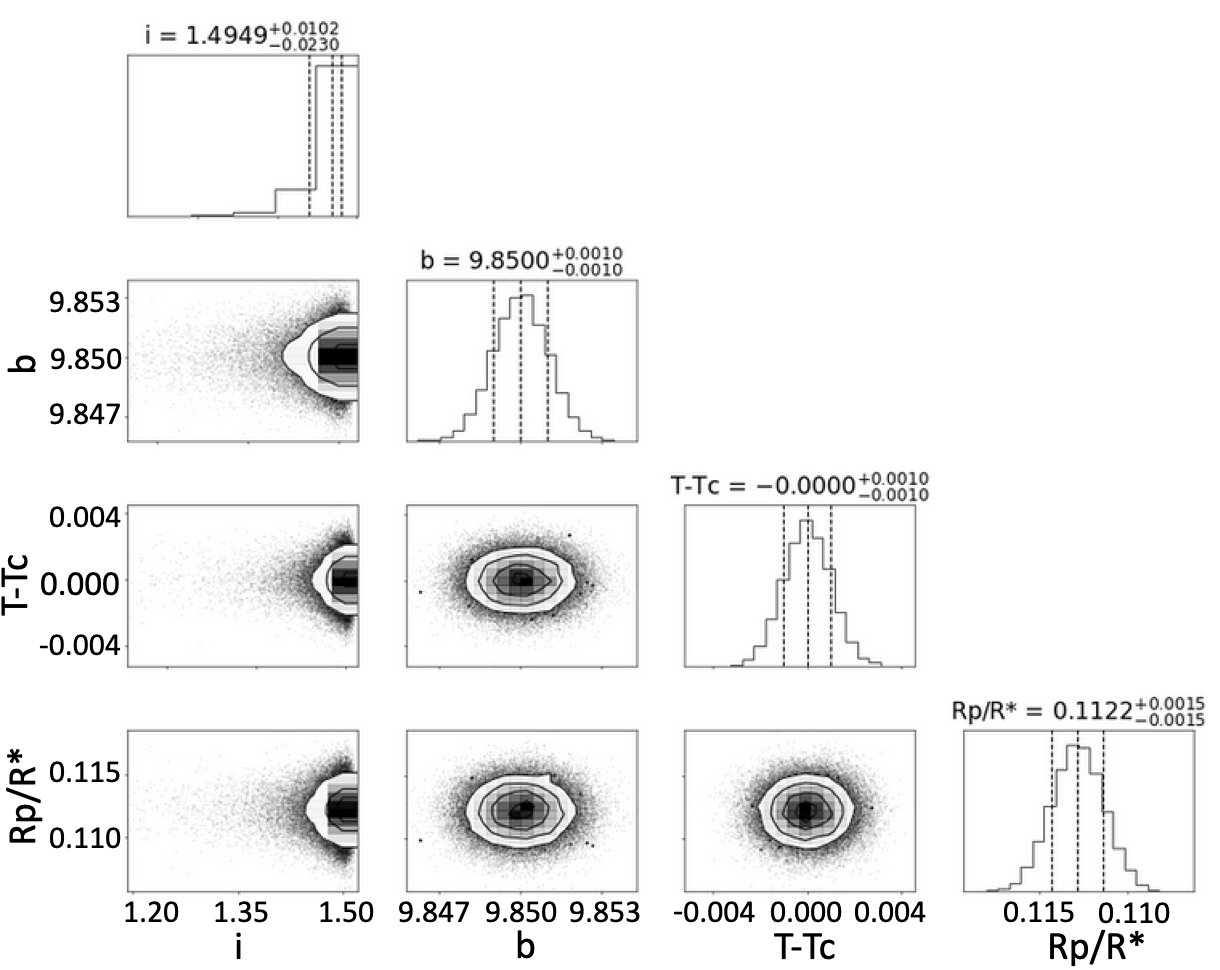} 
\caption{Posterior distribution of orbital inclination (i) in radians, impact parameter (b), the difference in mid-transit time ($T-T_{c}$) with $T_{C} = 2459079.932$ \citep{Nikolov_2014}, and the planet to star radius ratio ($R_{p}/R_{*}$) from the best-fit white lightcurve for HAT-P-1b.}
\label{posterior_hatp1}
\end{figure}

%

\begin{figure}
\centering
\includegraphics[scale=0.5]{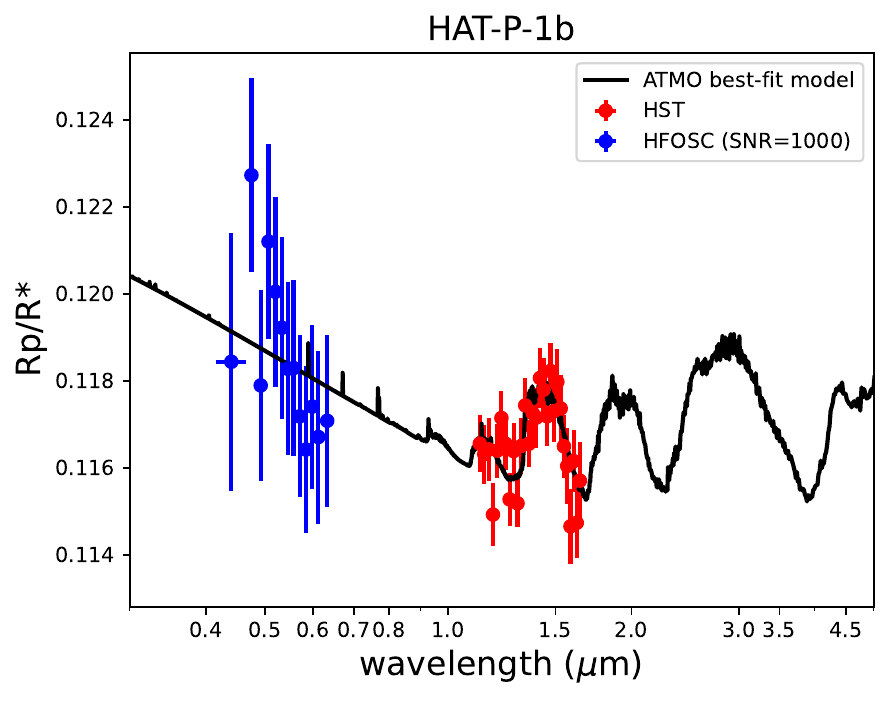} 
\caption{The best-fit ATMO model (black) for combined HFOSC (SNR=1000) (blue) and infrared observations from HST \citep{Wakeford_2013} (red) for the planet HAT-P-1b.}
\label{hatp1_hfosc_hst_model}
\end{figure}

\begin{figure}
\centering
\includegraphics[scale=0.5]{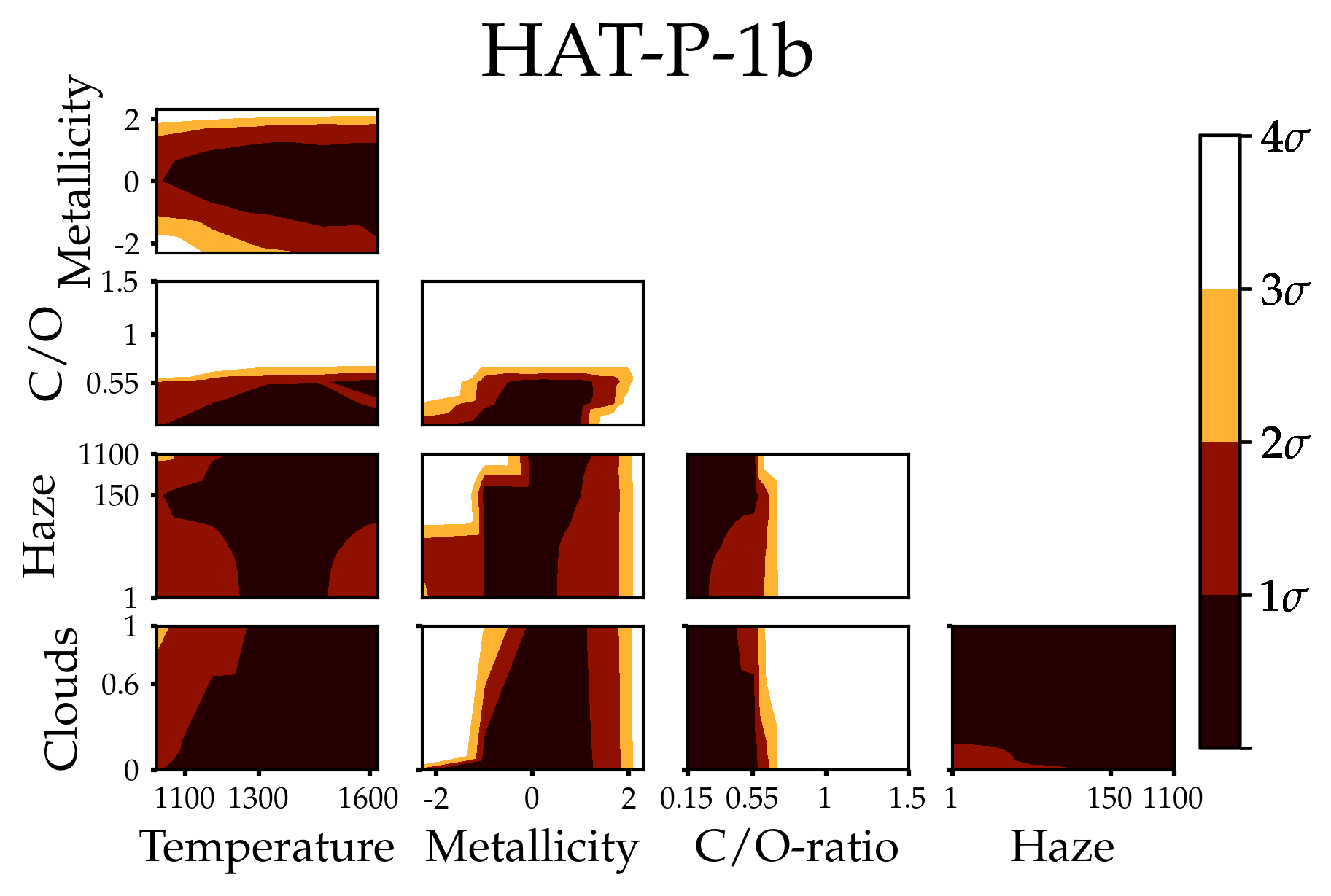} 
\caption{Figure showing $\chi^2$ map for HAT-P-1b when fitting model grid to IR observations from HST \citet{Wakeford_2013}. The contour, color maps and labels are same as Figure \ref{hatp1b_chimap_model}.}
\label{hatp1_corner_hst_model}
\end{figure}

\begin{figure}
\centering
\includegraphics[scale=0.5]{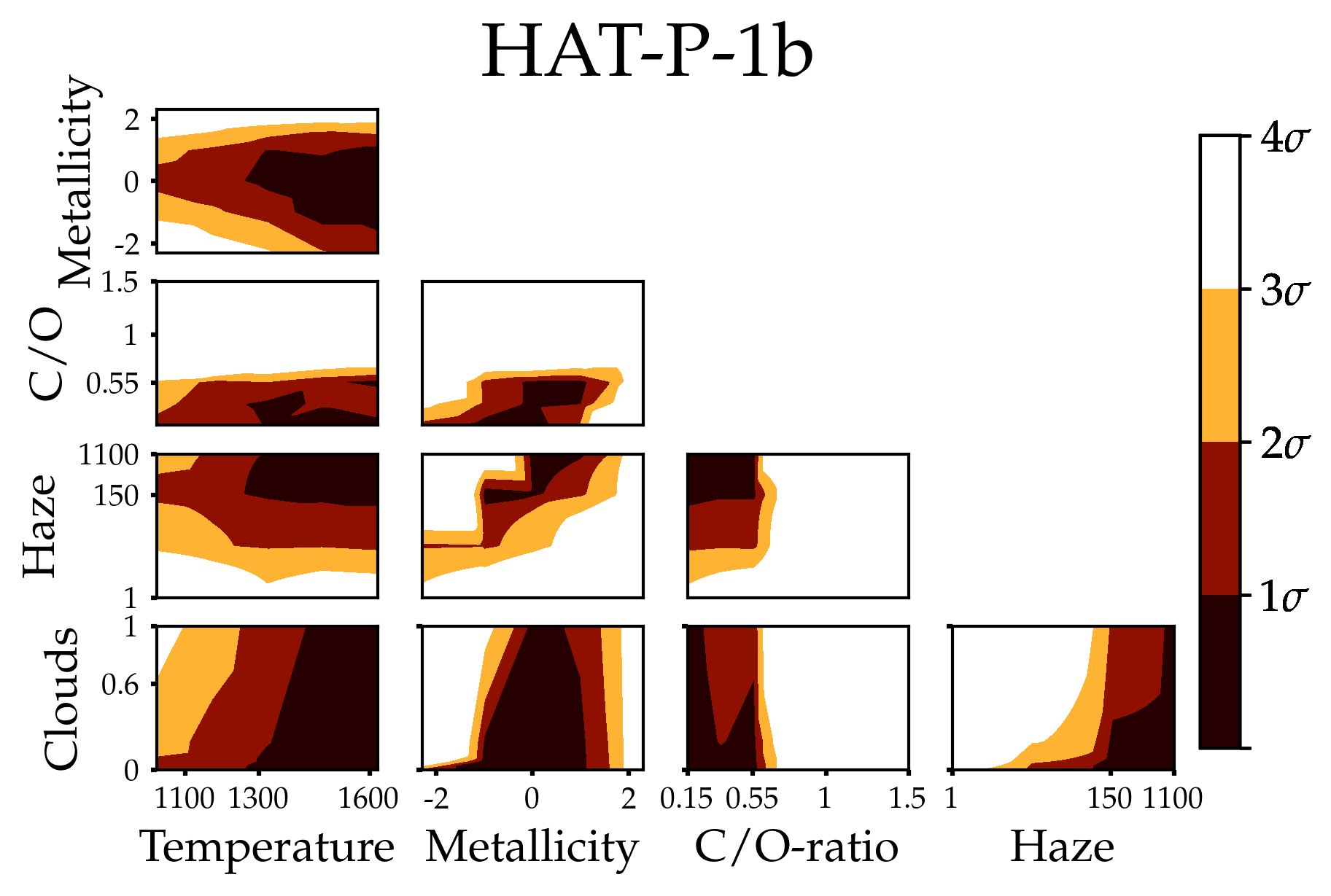} 
\caption{Figure showing $\chi^2$ map for HAT-P-1b when fitting model grid to HFSOC observations from this work and IR observations from HST \citet{Wakeford_2013}. The contour, color maps and labels are same as Figure \ref{hatp1b_chimap_model}.}

\label{hatp1_corner_hfosc_hst_model}
\end{figure}

\begin{figure}
\centering
\includegraphics[scale=0.5]{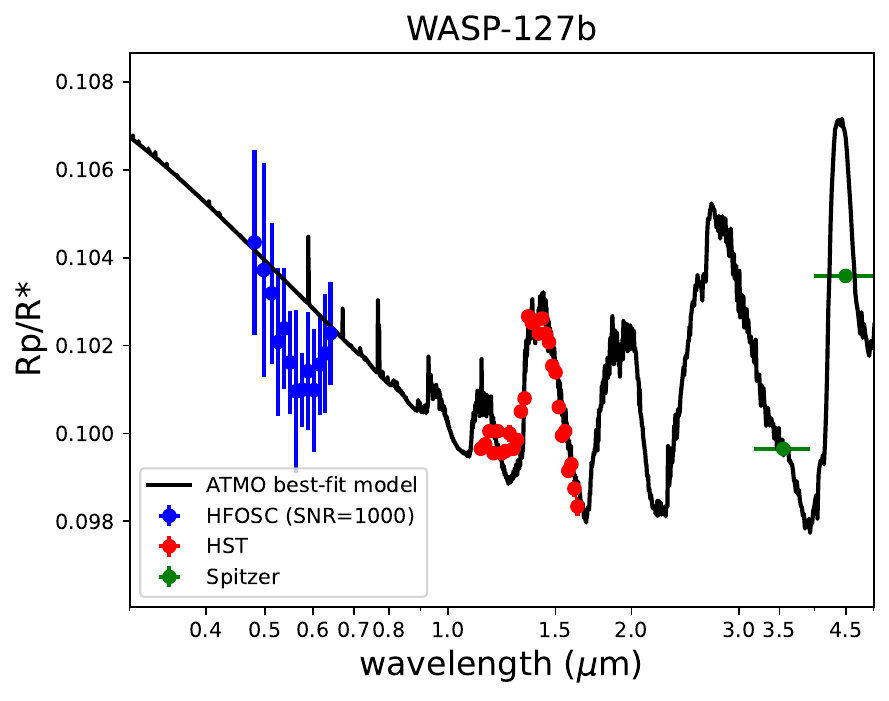} 
\caption{The best-fit ATMO model (black) for combined HFOSC (SNR=1000) (blue) and HST in IR \citep{Spake_2021} (red and green) for the planet WASP-127b.}
\label{wasp127_hfosc_hst_model}
\end{figure}

\begin{figure}
\centering
\includegraphics[scale=0.5]{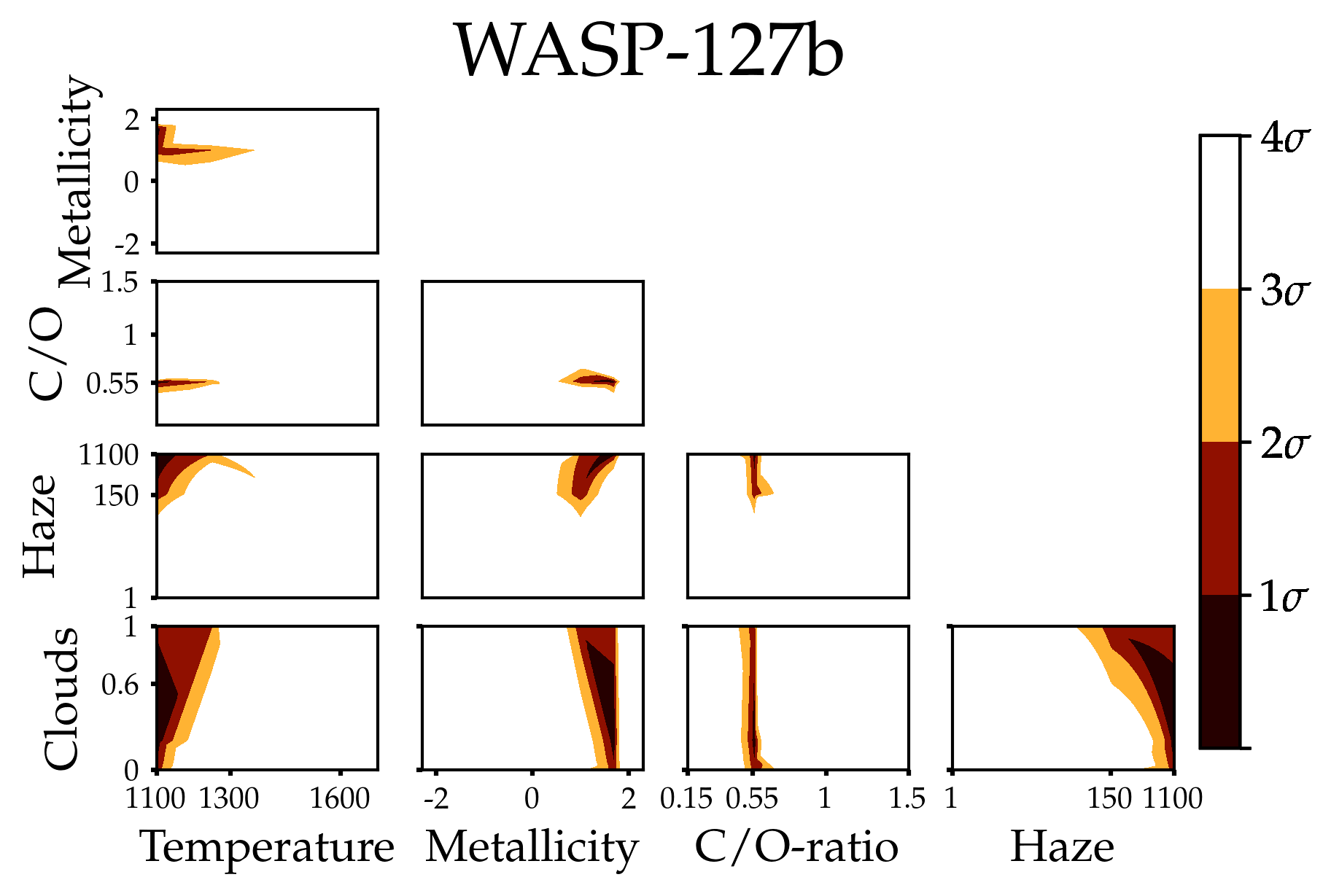} 
\caption{Figure showing $\chi^2$ map for WASP-127b when fitting model grid to HST and SPITZER observations from \citet{Spake_2021} in IR. The contour, color maps and labels are same as Figure \ref{hatp1b_chimap_model}.}
\label{wasp127_corner_hst_model}
\end{figure}

\begin{figure}
\centering
\includegraphics[scale=0.5]{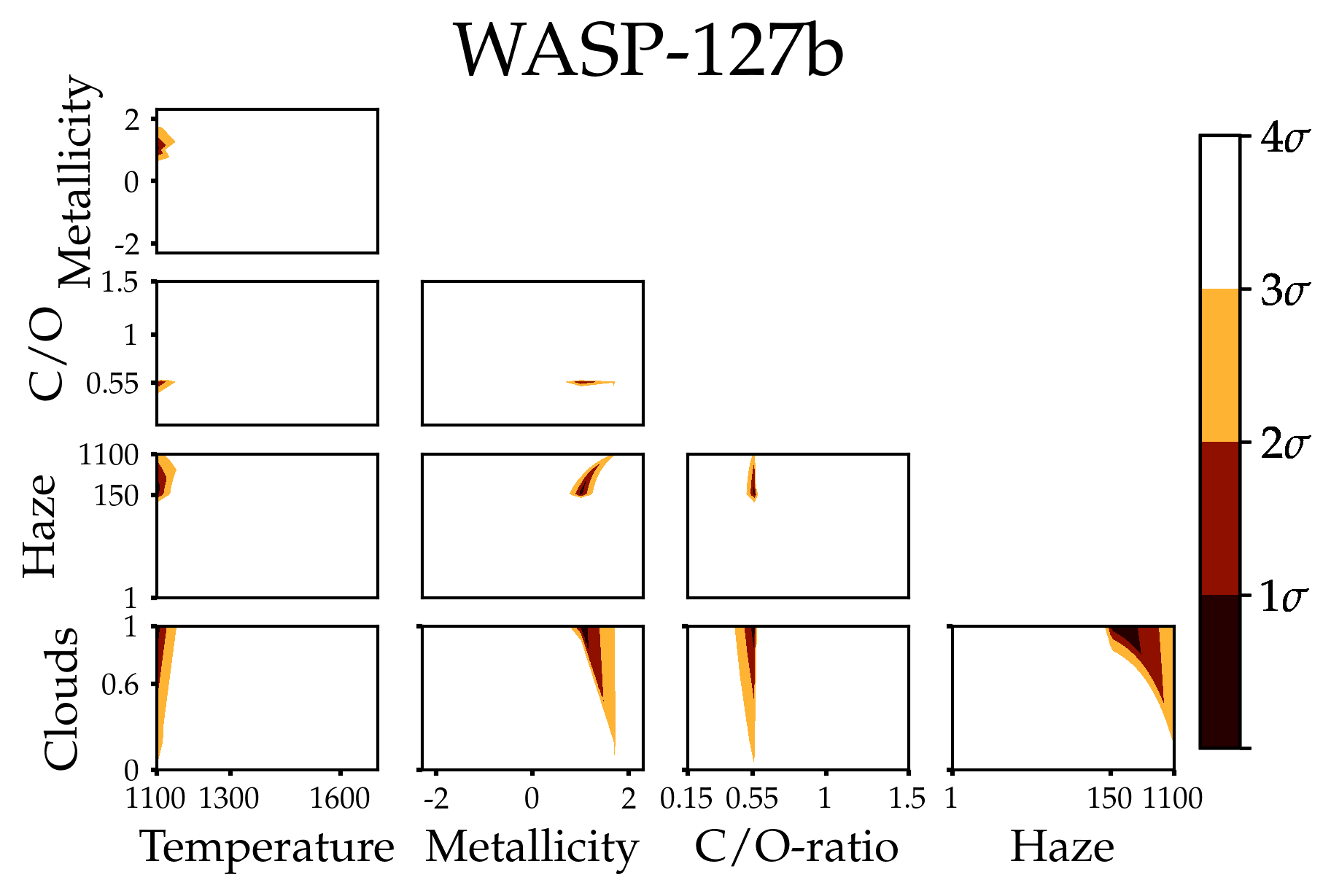} 
\caption{Figure showing $\chi^2$ map for HAT-P-1b when fitting model grid to HFSOC observations from this work and HST plus SPITZER observations from \citet{Spake_2021} in IR. The contour, color maps and labels are same as Figure \ref{hatp1b_chimap_model}.}
\label{wasp127_corner_hfosc_hst_model}
\end{figure}

\begin{figure}
\centering
\includegraphics[scale=0.55]{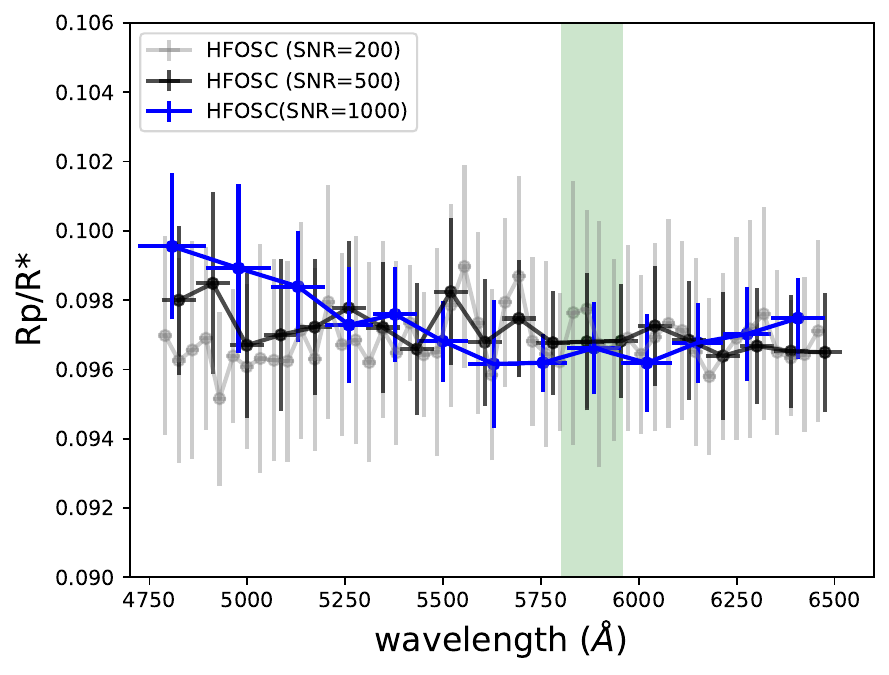} 
\caption{The observed transit spectra of WASP-127b in different wavelength bins SNR=1000 (blue), SNR=500 (black), and SNR=200 (grey). The vertical region in cyan shows the expected sodium doublet region. The Rayleigh scattering slope is not evident in both the transit spectra of SNR=200 and SNR=500. }
\label{wasp127_3bin}
\end{figure}

\bsp	
\label{lastpage}
\end{document}